\documentclass[a4paper,superscriptaddress,showpacs,twocolumn,floatfix,aps,prx]{revtex4-1}

\usepackage{amsmath,amsfonts,amssymb,amsxtra,mathdots,bm,xcolor,graphicx}
\usepackage{mathtools}
\usepackage{epsfig,grffile,tipa}
\usepackage[inner,final]{showlabels}
\usepackage{mathrsfs}
\usepackage{xspace}
\usepackage{afterpage}
\usepackage{hyperref}
\usepackage{lipsum}
\usepackage{soul}

\renewcommand{\vec}[1]{\bm{#1}}
\renewcommand{\tensor}[1]{\vec{\mathsf{#1}}}
\newcommand*{\stensor}[1]{\mathsf{#1}}

\newcommand*{\df}[1]{\text{d}#1\,}
\newcommand*{\vdf}[1]{\text{d}\vec{#1}\,}

\newcommand*{\dd}[2]{\frac{\df{#1}}{\df{#2}}}
\newcommand*{\kb}{k_{\textup{B}}}
\newcommand*{\kbt}{\kb T}
\newcommand*{\transp}[2][-2mu]{\ensuremath{\mskip1mu\prescript{\smash{\mathrm t\mkern#1}}{}{\mathstrut#2}}}%
\newcommand*{\adim}[1]{\widetilde{#1}}
\newcommand*{\De}{\textup{De}}
\newcommand*{\Rey}{\textup{Re}}
\newcommand*{\Cs}{\textup{Cs}}

\newcommand*{\test}[1]{#1_\star}

\newcommand*{\gp}{\mathcal{N}}
\newcommand*{\inv}[1]{{#1}^{-1}}
\newcommand*{\primed}[1]{{{#1}^{\prime}}}
\newcommand*{\grad}{\vec{\nabla}}
\newcommand*{\psum}{\sideset{}{'}\sum}
\newcommand*{\sumsph}[1]{\psum_{#1}\frac{m_{#1}}{\rho_{#1}}}
\newcommand*{\supt}[2]{{#1}^{\text{#2}}}
\newcommand*{\subt}[2]{{#1}_{\text{#2}}}

\newcommand*{\micro}[1]{\supt{#1}{(m)}}
\newcommand*{\macro}[1]{\supt{#1}{(M)}}
\newcommand*{\fdt}{\Delta t^{(c)}}
\newcommand*{\scaling}[1]{\mathcal{O}\left(#1\right)}

\DeclarePairedDelimiter\avg{\langle}{\rangle}
\DeclarePairedDelimiter\abs{\lvert}{\rvert}
\DeclarePairedDelimiter\norm{\lVert}{\rVert}
\DeclarePairedDelimiter\paren{(}{)}
\DeclarePairedDelimiter\bracket{[}{]}

\DeclarePairedDelimiterX\pcond[2]{(}{)}{#1\,\delimsize\vert\,\mathopen{}#2}

\DeclareMathOperator*{\argmax}{arg\,max}

\DeclareMathOperator*{\trace}{Tr}

\DeclareGraphicsExtensions{{.png,.pdf,.eps}}
\graphicspath{{./}}
\begin{document}
\onecolumngrid
Copyright © 2020 by American Physical Society. All rights reserved.

Published in : \href{https://journals.aps.org/prresearch/abstract/10.1103/PhysRevResearch.2.033107}{Physical Review Research \textbf{2}, 033107 (2020)}
\twocolumngrid
\title{Learning the constitutive relation of polymeric flows with memory}
\author{Naoki Seryo}
\author{Takeshi Sato}
\thanks{Present address: Institute for Chemical Research, Kyoto University,
  Kyoto 611-0011, Japan.}
\author{John J. Molina}
\thanks{Corresponding authors.\\
  \href{mailto:john@cheme.kyoto-u.ac.jp}{john@cheme.kyoto-u.ac.jp} (J.~J. Molina)\\
  \href{mailto:taniguch@cheme.kyoto-u.ac.jp}{taniguch@cheme.kyoto-u.ac.jp} (T. Taniguchi)}
\author{Takashi Taniguchi}
\thanks{Corresponding authors.\\
  \href{mailto:john@cheme.kyoto-u.ac.jp}{john@cheme.kyoto-u.ac.jp} (J.~J. Molina)\\
  \href{mailto:taniguch@cheme.kyoto-u.ac.jp}{taniguch@cheme.kyoto-u.ac.jp} (T. Taniguchi)}
\affiliation{Department of Chemical Engineering, Kyoto University,
  Kyoto 615-8510, Japan}
\date{\today}
\begin{abstract}
  We develop a learning strategy to infer the constitutive relation for the stress of polymeric flows with memory. We make no assumptions regarding the functional form of the constitutive relations, except that they should be expressible in differential form as a function of the local stress- and strain-rate tensors. In particular, we use a Gaussian Process regression to infer the constitutive relations from stress trajectories generated from small-scale (fixed strain-rate) microscopic polymer simulations. For simplicity, a Hookean dumbbell representation is used as a microscopic model, but the method itself can be generalized to incorporate more realistic descriptions. The learned constitutive relation is then used to perform macroscopic flow simulations, allowing us to update the stress distribution in the fluid in a manner that accounts for the microscopic polymer dynamics. The results using the learned constitutive relation are in excellent agreement with full Multi-Scale Simulations, which directly couple micro/macro degrees of freedom, as well as the exact analytical solution given by the Maxwell constitutive relation. We are able to fully capture the history dependence of the flow, as well as the elastic effects in the fluid. We expect the proposed learning/simulation approach to be used not only to study the  dynamics of entangled polymer flows, but also for the complex dynamics of other Soft Matter systems, which possess a similar hierarchy of length- and time-scales.
\end{abstract}
\pacs{}
\keywords{ Multi-Scale Simulations, Constitutive Relation, Learning, Gaussian Processes, Polymer Flows}
\maketitle
\section{Introduction}\label{s:intro}
Polymeric materials are ubiquitous in our modern industrial societies, having transformed our food, infrastructure, and modes of transportation. Not surprisingly, the 20th century has been dubbed the ``Polymer Age'' by Rubinstein and Colby\cite{Rubinstein2003}. There is a growing demand for producing more high-functioning polymeric products, and to do this in a cost-effective way. Unfortunately, there is still much we do not understood about polymer physics, particularly with regards to the fabrication method of sophisticated polymer products. At present, one of the preferred manufacturing methods for polymeric materials is polymer melt processing, where a molten polymer is extruded or molded into the desired shape, before allowing it to cool and solidify\cite{NationalResearchCouncil1994}. To accomplish these requirements, we need to understand not only the macroscopic flow behavior of the polymer process, but also the microscopic dynamics of the polymer chains, in order to reliably control the resultant properties of the product. However, it is not easy to understand such properties using only experimental observations, due to the hierarchy of length- and time-scales needed to characterize the microscopic polymer dynamics and the macroscopic flow. Computer simulations, which provide an alternative and complimentary approach, have become an indispensable tool for studying such systems. Molecular Dynamics (MD) simulations provide access to the detailed dynamics of polymer chains, but they are unable to deal with the macroscopic flow, because of the prohibitive cost. Computational Fluid Dynamics (CFD) simulations, with an appropriate constitutive equation for the stress tensor of the polymeric material\cite{Larson1988}, can predict the macroscopic flow behavior, but cannot provide any microscopic information on the constituent polymer chains. To address this issue, and elucidate the microscopic origin of the flow problems at hand, Multi-Scale Simulation (MSS) methods, which make it possible to simultaneously consider the dynamics at both scales, have been extensively developed over the last twenty years. Within the MSS approach, the macroscopic and microscopic degrees of freedom are coupled through the stress and strain-rate tensor fields. Originally proposed by Laso and \"Ottinger in 1993\cite{Laso1993,Ottinger2005}, with the so-called CONNFFESSIT model, these types of approaches represent the state-of-the-art in polymer rheology, as they provide a rigorous connection between the microscopic and macroscopic degrees of freedom\cite{Borodin2005,Yasuda2008, Murashima2010,Murashima2011,Masubuchi2012,Yasuda2014,Wu2018,Sato2017,Sato2017a,Sato2019a, Mu2019}. However, given the computational complexity, they have been limited to simple flow geometries and small system sizes, and have yet to be widely adopted within industry.

In this paper, we demonstrate how to leverage the power of Machine-Learning techniques to accelerate the MSS to the point where they are competitive with existing macroscopic descriptions, without any significant loss of accuracy. In particular, we will show that it is possible to learn the constitutive relation from training data generated from small system size microscopic simulations. To this end, we adopt a simple microscopic description, which models the polymers as non-interacting Hookean dumbbells. We note that the proposed method is applicable to any microscopic polymer model, the Hookean dumbbell model has been chosen for its simplicity. As is well known, in the limit when the number of dumbbells goes to infinity, the time evolution equation for the stress of such an ensemble converges to an analytic expression. This makes it possible to give a stringent assessment of our proposed ML approach.  We then perform simulations at fixed strain-rates, in order to measure the time-evolution of the stress. This information is used as the training data for a Gaussian Process (GP) regression, in order to learn the corresponding constitutive relation. A GP approach avoids over-fitting of the data and allows us to incorporate unknown and/or noisy data within a Bayesian framework, in a convenient and efficient manner\cite{Jaynes2003,Sivia2006,Rasmussen2005,Murphy2012}. The learned constitutive relations are then used in macroscopic flow simulations, opening the possibility of probing length- and time-scales that would be unreachable with standard MSS techniques. Previous work by Zhao et al. has used a similar approach to learn the constitutive relation of generalized Newtonian fluids\cite{Zhao2018}, but as proposed, the method cannot be applied to non-Newtonian viscoelastic materials that exhibit a history dependent flow. A recent extension of this method has used GP to learn the effective viscosity and relaxation time needed to parameterize a given viscoelastic constitutive relation\cite{Zhao2020}. In both cases, however, the functional form of the constitutive equation is a fixed input of the method. Here we show that this restriction can be lifted, and that the constitutive relation itself can be learned.

Compared to full MSS (using Hookean dumbbells), we obtain speedups of around two orders of magnitude, and we expect this will only increase when more realistic (computationally expensive) polymer models are used, as the cost of performing the macroscopic flow simulations remains constant. Finally, we note that the proposed learning strategy, which is here used to learn the constitutive relation of polymeric flows from microscopic data, is not limited only to polymeric materials. In fact, we envision similar approaches being used to bridge between length- and time-scales in other Soft-Matter systems, such as colloidal dispersions or cellular tissues.

\section{Multi-Scale Simulations}\label{s:mss}
\begin{figure}[ht!]
  \includegraphics[width=\columnwidth]{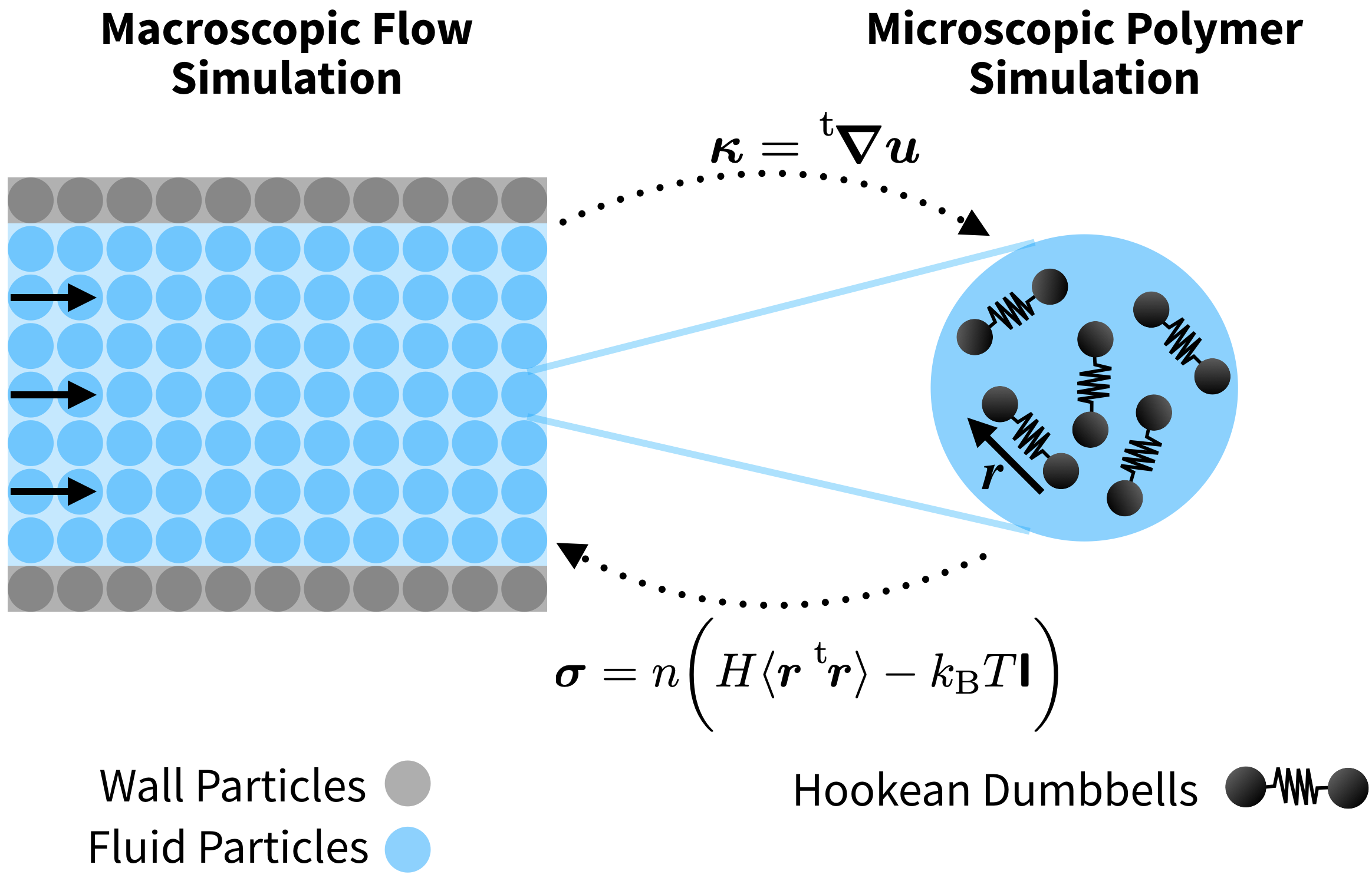}
  \caption{\label{f:mss}Schematic representation of the MSS method used to study the micro/macro coupling of polymeric flows. The fluid is discretized into fluid particles carrying mass and momentum, as well as microscopic polymer simulators. Solving for the dynamics of the polymers, under the macroscopically obtained velocity gradient $\tensor{\kappa}$, allows us to compute the microscopic polymer contribution to the local stress $\tensor{\sigma}$. The resultant stress distribution is then used to solve for the macroscopic flow dynamics.}
\end{figure}

\subsection{Macroscopic Model}
In order to consider the memory effects inherent to polymer flows we adopt a Lagrangian particle description to describe the dynamics of the fluid. This allows us to account for the convection of the polymer chains and the corresponding strain-rate history dependence on their dynamics. In particular, we will use the Smooth Particle Hydrodynamics (SPH) method(see Appendix~\ref{s:app_sph} for details)\cite{Monaghan1992}. The system is discretized into fluid particles, carrying mass, momentum, and  all relevant hydrodynamic variable for the problem under consideration.

Let $\vec{x}_i$ and $\vec{v}_i$ be the position and velocity of particle $i$; its time-evolution is determined by the following equations
\begin{align}
  \dd{\vec{x}_i}{t} & = \vec{v}_i \label{e:sph_x}                                                                                              \\
  \dd{\vec{v}_i}{t} & = \inv{\rho_i}\bracket{\grad\cdot\paren*{\tensor{\sigma} - p\tensor{I}}}_{\vec{x}_i} + \vec{F}(\vec{x}_i)\label{e:sph_v}
\end{align}
where $\rho_i$ is the density of the fluid particle and $\vec{F}(\vec{x}_i)$ is any external force acting on the particle (at position $\vec{x}_i$). The first term on the right-hand side of Eq.~\eqref{e:sph_v} corresponds to the forces on the particle due to internal stresses in the fluid (with $p$ the pressure field). The stress $\tensor{\sigma}$ comes from the time-dependent  state of the polymer chains, i.e., the orientation and stretching of the dumbbells.
The pressure $p_i$ is defined via the following quasi-incompressible equation of state
\begin{align}
  p_i & = \frac{{c_s}^2\rho_0}{\gamma}\left[\left(\frac{\rho_i}{\rho_0}\right)^\gamma-1\right]\label{e:sph_eos}
\end{align}
with $c_s$ the speed of sound and $\rho_0$ the initial density. Since we are interested in low-Reynolds number flows, we use $\gamma = 1$\cite{Morris1997}.

All that remains is to specify how $\tensor{\sigma}$ is computed. The simplest approach would be to adopt a constitutive equation (e.g., Oldroyd-B), instead, within a MSS approach, we place microscopic polymer simulators inside each fluid particle, in order to directly couple the microscopic and macroscopic degrees of freedom\cite{Murashima2010,Murashima2011,Sato2017,Sato2017a,Sato2019a}. The choice is then reduced to that of defining an appropriate microscopic model for the polymeric fluid.

\subsection{Microscopic Model}
To describe the rheological properties of the polymeric fluid, We choose the simplest possible microscopic model, that of non-interacting Hookean dumbbells. Thus, we place $N_p$ polymer chains inside each of the $N_f$ fluid particles, with each polymer chain represented by two point particles connected by a Hookean spring (Fig.~\ref{f:mss}). While more realistic microscopic models are known, such as the finite-extensible Hookean dumbbell model\cite{Bird1987a}, the Rouse model\cite{Rouse1953}, the Kremer-Grest beads-spring model\cite{Kremer1990}, the Doi-Edwards reptation model\cite{Doi1986}, and the slip-link models\cite{Masubuchi2001,Doi2003,Schieber2003,Likhtman2005,Uneyama2012}, the basic Hookean dumbbell model offers one main advantage over the others: In the limit when the number of dumbbells goes to infinity, the exact analytical constitutive equation for the stress is known and corresponds to that of a Maxwell viscoelastic fluid. This provides us with analytical results against which we can test our proposed learning strategy. However, the learning method we propose is not contingent on this choice, and in fact, will be most useful when considering more sophisticated polymer models, for which full MSS can become prohibitively expensive.

Since we do not consider interactions between dumbbells, the configuration of the system can be described solely in terms of the distance vector $\vec{r}$ between the two beads composing the dumbbell. The dynamics of the chains are then determined by the following Langevin equation for~$\vec{r}$\cite{Bird1987, Bird1987a}
\begin{align}
  \dd{\vec{r}}{t} & = \tensor{\kappa}\cdot\vec{r} - \frac{2}{\zeta} H \vec{r} + \frac{1}{\zeta}\vec{\xi}\label{e:dumbbell0} \\
  \tensor{\kappa} & = \transp{\tensor{\nabla}\vec{v}}\label{e:kappa}
\end{align}
where $H$ is the spring constant, $\tensor{\kappa}$ the velocity gradient tensor, $\zeta$ the friction coefficient, and $\vec{\xi}$ a random-force satisfying the fluctuation-dissipation theorem, $\avg{\vec{\xi}} = \vec{0}$ and $\avg{\vec{\xi}(t)\vec{\xi}(t^\prime)} = 4\kbt\zeta\tensor{I} \delta(t - t^\prime)$, with $\tensor{I}$ the unit tensor. The polymer contribution to the stress tensor can be expressed in Kramers' form as\cite{Bird1985}
\begin{align}
  \tensor{\sigma} & = n\paren[\Big]{H\avg{\vec{r}\transp{\vec{r}}} - \kbt\,\tensor{I}}\label{e:kirkwood0}
\end{align}
where $n$ is the number density of dumbbells, $\kb$ the Boltzmann constant, and $T$ the temperature. From Eqs.~\eqref{e:dumbbell0}-\eqref{e:kirkwood0}, the following constitutive equation for the stress tensor can be derived
\begin{align}
  \dd{\tensor{\sigma}}{t} & =
  \tensor{\sigma}\cdot\transp{\tensor{\kappa}} +
  \tensor{\kappa}\cdot\transp{\tensor{\sigma}} +
  n\kbt\paren[\Big]{\tensor{\kappa} + \transp{\tensor{\kappa}}} - \frac{4 H}{\zeta}\tensor{\sigma}\label{e:maxwell0}
\end{align}
which corresponds to the upper-convected Maxwell model.
\subsection{Nondimensionalized Equations}
To facilitate the simulation and analysis, we will rewrite the previous equations in non-dimensional form, introducing basic units for both macroscopic and microscopic descriptions. These two sets are labeled with an uppercase (M) and lowercase (m) superscript, for macroscopic and microscopic units, respectively.  At the macroscopic level, the characteristic scales are set by the length $L$ and velocity $U$ of the flow problem under consideration, and the corresponding fluid-advection time scale $\macro{\tau} = L/U$. Together with the shear-viscosity of the fluid $\eta_s$, and the initial (target) density $\rho_0$, we use these characteristic scales to set the units of density, time, length, and stress to be $\macro{\rho_0} = \rho_0$, $\macro{t_0} = \macro{\tau}$, $\macro{l_0} = L$, and $\macro{\sigma_0} = \eta_s/\macro{\tau} = \eta_s U/L$, respectively.  With these units, the equations governing the macroscopic dynamics, Eqs.\eqref{e:sph_x}-\eqref{e:sph_eos}, become
\begin{align}
  \dd{\adim{\vec{x}}_i}{\adim{t}} &=  \adim{\vec{v}}_i \label{e:asph_x}\\
  \Rey \dd{\adim{\vec{v}}_i}{\adim{t}} &= \inv{\adim{\rho}_i}\big[\adim{\grad}\cdot\left(\adim{\tensor{\sigma}} - \adim{p}\,\tensor{I}\right)\big]_{\adim{\vec{x}}_i} + \adim{\vec{F}}(\adim{\vec{x}}_i) \label{e:asph_v}\\
  \adim{p}_i &= \Cs^2\left(\adim{\rho} - 1\right)\label{e:asph_eos}
\end{align}
where a tilde ($\tilde{\cdot}$) denotes an adimensional variable. Here, the control parameters are the Reynolds number, defined as $\Rey = \rho_0 U L / \eta_s = \rho_0 U^2 / \macro{\sigma_0}$, and the dimensionless artificial sound speed $\Cs^2 = c_s^2 \macro{\rho_0} \macro{t_0} / \eta_s$.

At the microscopic level, the characteristic scales are given by the equilibrium length of the dumbbells $l_{\textup{eq}} = \sqrt{3\kbt/H}$, the dumbbell relaxation time $\micro{\tau} = \lambda = \zeta / 4 H$, and the shear viscosity of the Maxwell fluid $\eta_s \equiv n \kbt \lambda$. To facilitate comparisons between the microscopic and macroscopic models, we will use the same time and stress units for both, $\micro{t_0} = \macro{t_0}$ and $\micro{\sigma_0} = \macro{\sigma_0} = n \kbt \micro{\tau} / \macro{\tau}$, while the microscopic unit of length is taken to be the equilibrium dumbbell length $\micro{l_0} = l_{\textup{eq}}$. Thanks to the coupling between the macroscopic flow and the microscopic chain dynamics, the Deborah number, defined as the ratio of time-scales associated to the microscopic dumbbell relaxation and macroscopic fluid advection, $\De = \micro{\tau}/\macro{\tau}$, has appeared in the definition of the unit stress, $\micro{\sigma}=\macro{\sigma}=n\kbt\, \De$. With these microscopic units, Eqs.~\eqref{e:dumbbell0}-\eqref{e:maxwell0} become
\begin{align}
  \dd{\adim{\vec{r}}}{\adim{t}}         & =
  \adim{\tensor{\kappa}}\cdot\adim{\vec{r}} -
  \frac{1}{2\De}\adim{\vec{r}} +
  \sqrt{\frac{1}{3\De}}\adim{\vec{\xi}} \label{e:dumbbell}                                                                                                                                                                                                                                                       \\
  \adim{\tensor{\sigma}}                & = \frac{3}{\De}\paren*{\avg{\adim{\vec{r}}\transp{\adim{\vec{r}}}\,} - \frac{1}{3}\tensor{I}} \label{e:kirkwood}                                                                                                                                                                \\
  \dd{\adim{\tensor{\sigma}}}{\adim{t}} & = \adim{\tensor{\sigma}}\cdot\transp{\adim{\tensor{\kappa}}} + \adim{\tensor{\kappa}}\cdot\transp{\adim{\tensor{\sigma}}} + \frac{1}{\De}\paren[\Big]{\adim{\tensor{\kappa}} + \transp{\adim{\tensor{\kappa}}}} - \frac{1}{\De}\adim{\tensor{\sigma}}\label{e:maxwell}
\end{align}
with the Deborah number $\De$ as the only control parameter. Note that $\adim{\tensor{\kappa}} = \tensor{\kappa} \micro{t_0}$ and $\adim{\vec{\xi}} = \sqrt{\micro{t_0}/\paren*{4\kbt\zeta}}\vec{\xi}$, with $\avg{\adim{\vec{\xi}}\,} = \tensor{0}$ and $\avg*{\adim{\vec{\xi}}\paren*{\adim{t}\,} \adim{\vec{\xi}}\paren*{\adim{t}^\prime}} = \tensor{I}\delta(\adim{t}-\adim{t}^\prime)$. Finally, we stress that, while we have chosen $\macro{\tau}$ as unit of time, it can also be useful to use the dumbbell relaxation time $\micro{\tau}\equiv \lambda$ as reference. Scaled time values using both units are directly related by the Deborah number, with $\adim{t} = t/\macro{\tau} = (t /\lambda)\De$.

In summary, the MSS method amounts to solving Eqs.~\eqref{e:asph_x}-\eqref{e:asph_eos} and \eqref{e:dumbbell}-\eqref{e:kirkwood}. 
  With the Reynolds number $\Rey$, the Deborah number $\De$, and the dimension-less artificial sound speed $\Cs$ as the main control parameters. Unless stated otherwise, we focus on the Deborah number regime $10^{-3}\le \De \le 10$, at low-Reynolds numbers $\Rey\lesssim 1$; with the sound-speed $\Cs=10$ chosen to keep the density variations within $\lesssim 3\%$\cite{Morris1997}, while still allowing for a relatively large simulation time step $\Delta \adim{t}\simeq 10^{-6}$. The strain-rate tensor is evaluated at the macroscopic level and used as input for the microscopic simulators, in order to evolve the configuration of the polymers (dumbbells). Then, the polymer contribution to the stress is computed and used as input to the macroscopic flow simulation, in order to update the fluid velocity, and the process is repeated (see Figure~\ref{f:mss}). This method has been successfully used to study a polymer melt-spinning process\cite{Sato2017a}, as well as flows of well-entangled polymer melts in a contraction-expansion channel (with the dumbbell model replaced by the Doi-Takimoto Slip-Link model)\cite{Sato2019a}. Unfortunately, while the predictive capabilities of such a bottom-up approach represent the current state-of-the-art in the field of polymer simulations, their heavy computational cost has mostly limited them to relatively simple flow-geometries in 2D. To address this issue, we propose a learning strategy based on Gaussian Processes, in order to uncouple the microscopic and macroscopic degrees of freedom in the governing equations. However, our method remains Multi-Scale, in the sense that the microscopic model is used to generate the training data used to learn the appropriate constitutive equation, i.e. how to evolve $\tensor{\sigma}$. This will allow us to consider the time-evolution of the stress at the macroscopic level, but in a way that satisfies the dynamics of the underlying microscopic polymer model.

\section{Learning Method}\label{s:learn}
\begin{figure*}[ht!]
  \centering
  \includegraphics[width=0.95\textwidth]{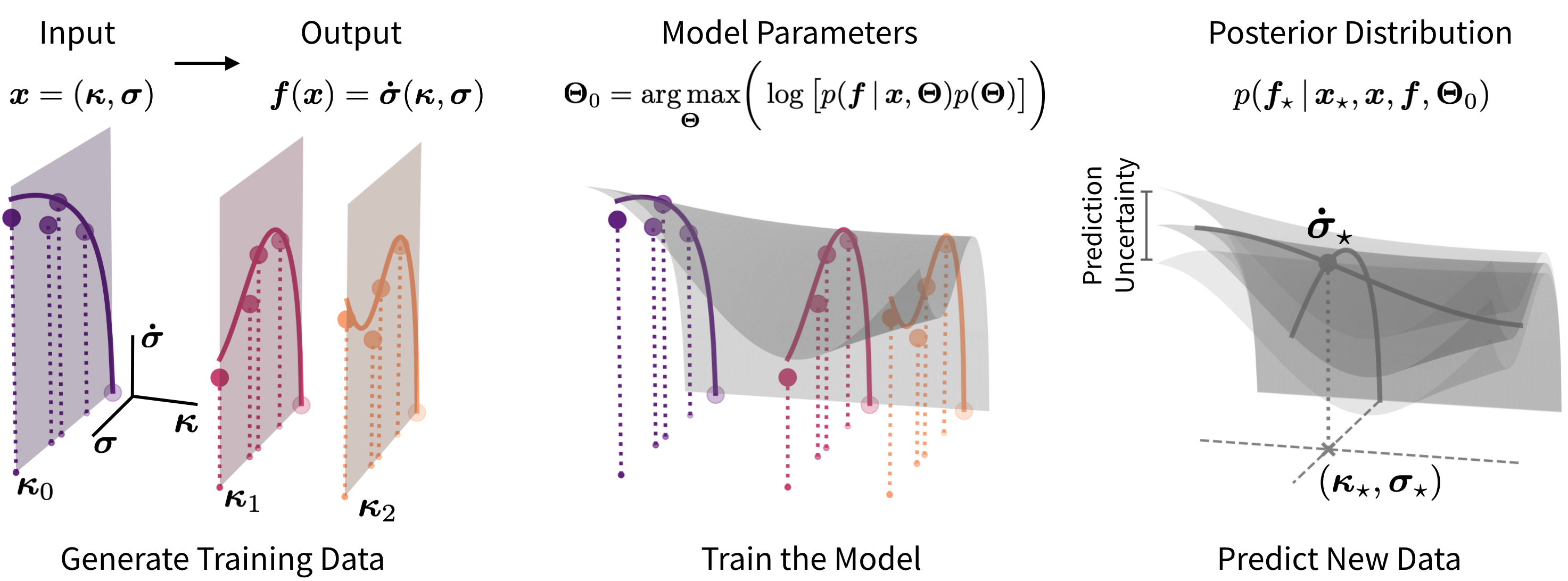}
  \caption{\label{f:gp}Schematic representation of the GP-MSS strategy used to learn the constitutive relation of polymeric flows from microscopic data. (1) We perform small-scale microscopic simulations at fixed strain-rate $\tensor{\kappa}$, in order to obtain the input $(\tensor{\kappa},\tensor{\sigma})$  and output $\dot{\tensor{\sigma}}$ training data. (2) Placing a GP prior on the constitutive relation, $\dot{\tensor{\sigma}}\sim\gp(\mu, K)$, the training data is used to learn the hyper-parameters $\vec{\Theta}$ (specifying the function variance and length-scales) of the GP, by maximizing the posterior distribution for $\vec{\Theta}$. (3) The probability distribution for the constitutive relation at new ``test'' inputs $(\test{\tensor{\kappa}}, \test{\tensor{\sigma}})$ is then specified by a conditional GP, i.e., $\left.\test{\dot{\tensor{\sigma}}}\right\lvert\tensor{\sigma}\sim \gp(\vec{\nu}, \Sigma)$. We take the average value to be the best estimate, $\test{\dot{\tensor{\sigma}}} = \vec{\nu}$, and use this within a macroscopic simulation, in order to update the stress at each point in the fluid. The prediction uncertainty is given by the corresponding covariance $\Sigma$, and it can be used to perform on-the-fly diagnostics\cite{Zhao2018}.}
\end{figure*}
\subsection{Gaussian Processes (GP)}
Formally, a Gaussian Process is defined as ``a collection of random variables, any finite number of which have a joint Gaussian distribution''\cite{Rasmussen2005}. It provides a (prior) probability distribution over functions $f$, allowing us to use known values of the function, the so-called ``training'' data, to infer the values of the function at new ``test" positions. Let $f(\vec{x})$ be a function from $\mathbb{R}^D$ to $\mathbb{R}$, which is sampled at $N$ values of the input $\vec{x}$, $\vec{x}_i\in\mathbb{R}^D (i=1,\ldots,N)$. We denote by $\vec{X} = (\vec{x}_1,\ldots,\vec{x}_N)$ the $D\times N$ design matrix, and $\vec{f}= \vec{f}(\vec{X}) = (f_1, \ldots, f_N)$ the corresponding output matrix, with $f_i = f(\vec{x}_i)$. The $f_i$ are considered to be correlated random variables, where the correlation between any two of them, $f_i$ and $f_j$, is assumed to be a function only of the input values, $\vec{x}_i$ and $\vec{x}_j$. The joint distribution for the $f_i$ is a multi-variate Gaussian, specified in terms of an average function $\mu(\vec{x})$ and a correlation function $k(\vec{x},\vec{x}^\prime)$.

The probability of observing function values $\vec{f}$ at $\vec{X}$, given $\mu$ and $k$, is
\begin{align}
  p & \pcond{\vec{f}(\vec{X})}{\vec{X}, \mu, k}= \frac{1}{\sqrt{\paren*{2\pi}^N\det{K(\vec{X},\vec{X})}}}\label{e:gp_multi}                  \\
    & \qquad\times\exp{\bracket[\Big]{-\frac{1}{2}\transp{\delta\vec{f}(\vec{X})}\cdot \inv{K(\vec{X},\vec{X})}\cdot\delta\vec{f}(\vec{X})}}
  \notag
\end{align}
where $\delta \vec{f} = \vec{f} - \vec{\mu}$, and $K(\vec{X},\vec{X})$ denotes the $N\times N$ correlation matrix, whose $(i,j)$-th entry is defined as $K(\vec{A},\vec{B})_{ij} = k(\vec{A}_i, \vec{B}_j)$. Note that a GP is uniquely defined in terms of its average and correlation functions,
\begin{align}
  \avg[\big]{f(\vec{x})}                                 & = \mu(\vec{x}) \label{e:gp_avg}             \\
  \avg[\big]{\delta f(\vec{x}) \delta f(\vec{x}^\prime)} & = k(\vec{x},\vec{x}^\prime)\label{e:gp_cov}
\end{align}
Without loss of generality,  and in the absence of any information about $f$, one can take $\mu(\vec{x}) = 0$, which leaves only $k$ to be specified. Note that no assumptions have been made regarding the functional form of $f$, the correlation function $k$ only determines the higher-order properties of the family of functions from which $f$ is drawn, such as continuity, differentiability, and periodicity. Following Rassmussen and Williams\cite{Rasmussen2005}, we also use the following shorthand notation to specify a GP
\begin{align}
  \vec{f} & \sim \gp\paren{\vec{\mu}, K} \label{e:gp}
\end{align}
which should be interpreted according to Eq.~\eqref{e:gp_multi}.
The fact that the function values at different positions ($f(\vec{x})$ and $f(\primed{\vec{x}})$) are correlated, with $\avg{\delta f(\vec{x}) \delta f(\primed{\vec{x}})} = k(\vec{x}, \primed{\vec{x}})$, is what allows us to make predictions. Basically, the data for $f(\vec{x})$, measured at the ``training" points $\vec{x}$, allows us to ``learn" how the data is correlated. This is done by inferring the hyper-parameters $\vec{\Theta}$ of the correlation function $k$, given the training data set. These hyper-parameters $\vec{\Theta}$ determine the precise shape of $k$, and thus the properties of the random functions $f$ drawn from the GP. Once we have learned how, and to what degree, the function values are correlated with each other, we use this information, together with the known values of $f$ at the training points $\vec{x}$, to predict the values of the function $\test{f}$ at new ``test" locations $\test{\vec{x}}$.

Assuming the $N$ input values consist of $n$ training points and $m$ test points, we partition the input and output data into training and test data sets, to arrive at the following (prior) joint distribution
\begin{align}
  \begin{bmatrix}
    \vec{f} \\
    \test{\vec{f}}
  \end{bmatrix} & \sim
  \gp\paren*{\begin{bmatrix}\vec{\mu}\paren*{\vec{X}} \\ \vec{\mu}\paren*{\test{\vec{X}}}\end{bmatrix},
    \begin{bmatrix}
      K(\vec{X}, \vec{X})        & K(\vec{X}, \test{\vec{X}})        \\
      K(\test{\vec{X}}, \vec{X}) & K(\test{\vec{X}}, \test{\vec{X}})
    \end{bmatrix}}\label{e:gp_joint}
\end{align}
where $\vec{X} = (\vec{x}_1, \ldots, \vec{x}_n)$ is now the design matrix for the training points and $\test{\vec{X}} = (\vec{x}_{1\star},\ldots,\vec{x}_{m\star})$ that of the test points. Thus, the correlation sub-matrices $K(\vec{X},\vec{X})$, $K(\vec{X},\test{\vec{X}}) = \transp{K(\test{\vec{X}}, \vec{X})}$, and $K(\test{\vec{X}}, \test{\vec{X}})$ have dimensions $n\times n$, $n\times m$, and $m\times m$, respectively. We stress the fact that Eqs.~\eqref{e:gp}~and~\eqref{e:gp_joint} are equivalent, at this point we have simply relabeled the points as belonging to either the training or test data sets. In addition, if the training data is not known exactly, but subject to noise, we can account for this by adding the corresponding contribution to the covariance sub-matrix. For example, in the presence of Gaussian noise with variance $\sigma^2$, we would simply replace $K(\vec{X},\vec{X})$ with $K(\vec{X},\vec{X}) + \sigma^2\tensor{I}_{n\times n}$.

The benefit of using GP comes from the Gaussian form of the distributions, since this allows us to perform most of the calculations analytically. In particular, the conditional distribution for the function values at the test points, conditioned on the training data, can be obtained from Bayes' rule as
$p\pcond{\test{\vec{f}}}{\vec{f}} = p(\test{\vec{f}},\vec{f})/p(\vec{f})$, and it results in yet another GP\cite{Rasmussen2005}
\begin{align}
            & \qquad\qquad\qquad\test{\vec{f}}\lvert \vec{f} \sim \gp\paren*{\vec{\nu}, \Sigma}  \label{e:gp_cond}                                 \\
  \vec{\nu} & = \vec{\mu}(\test{\vec{X}}) + K(\test{\vec{X}}, \vec{X})\cdot\inv{K(\vec{X}, \vec{X})}\cdot\delta \vec{f}(\vec{X}) \notag            \\
  \Sigma    & = K(\test{\vec{X}}, \test{\vec{X}}) - K(\test{\vec{X}}, \vec{X})\cdot \inv{K(\vec{X}, \vec{X})}\cdot K(\vec{X},\test{\vec{X}})\notag
\end{align}
The prediction for the function values at the test points is then given by the mean $\vec{\nu}$, with the covariance matrix $\Sigma$ providing a measure of the uncertainty at each point. Conceptually, one can interpret this prediction as the result of drawing random functions from the prior $\vec{f}\sim\gp(\vec{\mu},K)$, and keeping only those that are consistent with the measured training data. The average and variance of the functions that remain will coincide with $\vec{\nu}$ and $\Sigma$, i.e., $\test{\vec{f}}\sim\gp(\vec{\nu}, \Sigma)$.

We have used a squared-exponential kernel for all our GP regressions. For a 1D regression problem, the kernel is defined as
\begin{align}
  k(x, x^\prime;\Gamma, l)    & = \Gamma^2 k_{\text{SE}}(x,x^\prime;l)\label{e:gp_sqexp} \\
  k_{\text{SE}}(x,x^\prime;l) & = \exp{\left[-\frac{(x-x^\prime)^2}{2 l^2}\right]}\notag
\end{align}
Here, $\vec{\Theta} = (\Gamma,l)$ are the hyper-parameters that must be inferred from the data, with $\Gamma$ specifying the amplitude of the function variance and $l$ the characteristic length-scale over which the function is varying in the $x$-dimension. This commonly used Squared-Exponential kernel results in GPs that are infinitely differentiable, but other choices are possible, and might be more suitable for a given learning problem. Kernels for higher dimensional function spaces can be defined by taking sums or products of 1D kernels\cite{Rasmussen2005,Duvenaud2014}. As an example, the following are valid GP kernels for 2D functions, with input $\vec{x}=(x_1,x_2)$
\begin{align*}
  k_{\text{sum}}(\vec{x}, \vec{x}^\prime; \Gamma, l_1, l_2) & = \Gamma^{2}\big(k_{\text{SE}}(x_1, x_1^\prime;l_1) + k_{\text{SE}}(x_2, x_2^\prime;l_2) \big) \\%\label{e:gp_ksum}\\
  k_{\text{mul}}(\vec{x}, \vec{x}^\prime; \Gamma, l_1, l_2) & = \Gamma^{2} k_{\text{SE}}(x_1, x_1^\prime; l_1) \cdot\,k_{\text{SE}}(x_2, x_2^\prime; l_2)    %\label{e:gp_kmul}
\end{align*}
In principle, we have independent sets of hyper-parameters for each dimension, but for simplicity, when using additive kernels we will assume that the amplitude of the variance is equal along all dimensions, i.e., $\Gamma = \Gamma_1 = \Gamma_2$.

\subsection{GP Accelerated Multi-Scale Simulations}
The idea of learning a constitutive relation for polymer flows from microscopic data is not entirely new. Indeed, previous work by Zhao et al.\cite{Zhao2018} has considered this precise problem, in order to perform macroscopic polymer flow simulations without having to introduce an arbitrary constitutive relation. In this work, they have assumed a generalized Newtonian model, and used simple-shear simulations to learn the apparent viscosity $\supt{\eta}{(app)} = \sigma_{xy}/\dot{\gamma}$, of (inelastic) non-Newtonian fluids as a function of the shear-rate $\dot{\gamma}$. When performing the macroscopic flow simulations, the maximum shear rate (computed from the second invariant of the strain rate tensor) is taken as the local shear rate, and used within a GP regression scheme in order to obtain $\supt{\eta}{(app)}$, and thus the local shear stress $\sigma_{xy} = \supt{\eta}{(app)} \dot{\gamma}$. In practice, this amounts to placing a GP prior on the stress tensor itself
\begin{align}
  \tensor{\sigma}\sim \gp(\mu, K)\label{e:zhao_model}
\end{align}
However, as acknowledged by the authors, this excludes many interesting rheological properties of (viscoelastic) non-Newtonian fluids, as it does not allow for any type of history dependence in the flow, and relies on a separation of time-scales between the microscopic and macroscopic dynamics. This history dependence, which arises from the internal stresses, is one of the most significant features of polymeric flows. Subsequent work by Zhao et al.\cite{Zhao2020} has considered viscoelastic flows, but the learning is restricted to parameterizing a constitutive relation with a predetermined functional form. Therefore, the goal of the current work is to generalize the learning strategy, so that it can be used to model viscoelastic polymeric fluids without having to specify any constitutive relation.

Following Zhao et al.\cite{Zhao2018,Zhao2020}, we also use small-scale microscopic simulations of polymer chains at fixed strain-rates to generate the training data necessary to learn the constitutive relation. However, we now place a GP prior on the time-derivative of the stress-tensor
\begin{align}
  \frac{\df{}}{\df{t}}\tensor{\sigma}\equiv\dot{\tensor{\sigma}}\sim \gp(\mu, K)\label{e:ssmt_model}
\end{align}
not on the stress-tensor itself (or the effective viscosity). This difference allows us to consider the time-dependent memory effects crucial to describe the dynamics of polymer chains. Even in the absence of polymer entanglement, this memory effect is non-negligible, thanks to the finite relaxation time of the polymer stretching and reorientation. No assumptions are made regarding the form of the constitutive equation, except for the fact that it should be expressed in differential form, as a function of the local instantaneous stress $\tensor{\sigma}$ and strain-rate $\tensor{\kappa}$ tensors. This includes most commonly used differential models, such as the Maxwell, Jeffreys, and Oldroyd fluids\cite{Larson1988}. Finally, while it is possible to consider correlated output, we use a separate GP for each independent component of the stress tensor, resulting in $D(D+1)/2$ GP regressions in $D$-dimensions.

Here, we have assumed a separation of length-scales between the microscopic and macroscopic descriptions, such that at the microscopic level the system can be considered to be homogeneous, i.e., all field gradients are effectively zero. It is only at the macroscopic-level where field gradient induced phenomena are incorporated. The appropriate microscopic length scale needed for this approximation to be valid will depend on the precise flow problem being studied. For the systems presented here it is given by the characteristic size of the polymer chains. It is still possible (in principle) to apply the same type of learning strategy to microscopic models in which field-gradients are not homogeneous. In such cases, the constitutive relations would be functions of the local stresses, strains, and their gradients. However, such considerations lie outside the scope of the current work.

A schematic diagram of the three-step learning strategy, consisting of data generation, learning, and prediction steps, is given in Fig.~\ref{f:gp}. For the first step, we performed fixed strain-rate simulations of the 3D microscopic model, corresponding to either simple-shear or planar elongational flow
\begin{align}
  \tensor{\kappa}^{\text{(shear)}}        & = \begin{pmatrix}
    0 & \dot{\gamma} & 0 \\
    0 & 0            & 0 \\
    0 & 0            & 0
  \end{pmatrix}, \begin{pmatrix}
    0            & 0 & 0 \\
    \dot{\gamma} & 0 & 0 \\
    0            & 0 & 0
  \end{pmatrix} \label{e:k_shear} \\
  \tensor{\kappa}^{\text{(elongational)}} & = \begin{pmatrix}
    \dot{\varepsilon} & 0                   & 0 \\
    0                 & - \dot{\varepsilon} & 0 \\
    0                 & 0                   & 0
  \end{pmatrix}\label{e:k_elong}
\end{align}
with $\dot{\gamma}$ and $\dot{\varepsilon}$ the shear and elongational flow rates, respectively. These particular flows are chosen because the learned constitutive relations will be used in simulations with an imposed 2D flow, where the $z$-direction is assigned to be neutral. However, if one wants to apply this strategy to a more complex situation, additional applied flows, with different deformation rate tensors, will need to be used during the learning process. All simulations were started from a random and isotropic initial configuration of the polymer chains ($\avg{\tensor{\sigma}} = 0$). The simulations were performed until a steady-state was reached, such that $\dot{\tensor{\sigma}}\simeq 0$. Each simulation provides us with a trajectory $(t, \tensor{\sigma}(t))$, from which we can compute $\dot{\tensor{\sigma}}(t)$. We randomly chose a fraction of these tuples $(\tensor{\kappa}, \tensor{\sigma}, \dot{\tensor{\sigma}})$ to serve as training data for the GP regression, with input $\vec{x}=(\tensor{\kappa}, \tensor{\sigma})$ and output $f(\vec{x}) = \dot{\tensor{\sigma}}$. To compute $\dot{\tensor{\sigma}}$ we use a finite-difference approximation over a characteristic time-scale $\fdt$.
To reduce the noise in the measurements for $\tensor{\sigma}$ and/or $\dot{\tensor{\sigma}}$, it is recommended to apply a data smoothing operation, the details of which are given below.

The second step is to train the model, i.e. to determine the hyper-parameters $\vec{\Theta}$ of the kernel function $k(\vec{x},\vec{x}^\prime)$, as well as any unknown uncertainties in the test data for $\dot{\tensor{\sigma}}$. The posterior probability distribution for the hyper-parameters, given the model and measured training data, is determined from Bayes' theorem as
\begin{align}
  p\pcond{\vec{\Theta}}{\vec{X},\vec{f}} & \propto p\pcond{\vec{f}}{\vec{X},\vec{\Theta}} p(\vec{\Theta})\label{e:logp_theta}
\end{align}
where the likelihood $p\pcond{\vec{f}}{\vec{X}, \vec{\Theta}}$ is given by the corresponding GP, Eq.~\eqref{e:gp_multi}, and $p(\vec{\Theta})$ is a suitably chosen prior for the hyper-parameters.
A full Bayesian treatment, in which we integrate out the hyper-parameters and propagate the uncertainties during the simulation is possible, but for simplicity, we will consider only a point-wise solution. For the 1D problem considered below, we take the posterior average
\begin{align}
  \vec{\Theta}_0^{(\text{avg})} = \int\vdf{\Theta}\, \vec{\Theta}\, p\pcond{\vec{\Theta}}{\vec{X}, \vec{f}}\label{e:theta_avg}
\end{align} estimated from Hamiltonian Monte-Carlo (HMC) simulations. For the 2D problem, we instead use a stochastic gradient-based optimization method\cite{Carleo2019} to maximize the log-posterior, and find the ``optimal'' value
\begin{align}
  \vec{\Theta}_0^{(\text{map})} = \argmax_{\vec{\Theta}} \log{p\pcond{\vec{\Theta}}{\vec{X}, \vec{f}}}\label{e:theta_map}
\end{align}

Finally, the optimized $\vec{\Theta}_0$ are used to parameterize the conditional distributions for the test data $\test{\vec{f}}\lvert \vec{f}$, as defined in Eq.\eqref{e:gp_cond}. In practical terms, we use the (conditional) mean $\vec{\nu} = \avg{\test{\dot{\tensor{\sigma}}}} $ as our prediction for the constitutive relation, such that the stress of each fluid particle is updated according to
\begin{align}
  \tensor{\sigma}_i(t + \Delta t) & = \tensor{\sigma}_i(t) + \avg{\test{\dot{\tensor{\sigma}}}}_i \Delta t\label{e:gpmss_sigma}
\end{align}
where $\avg{\test{\dot{\tensor{\sigma}}}}_i$ is a function of the training data $\vec{x} = (\tensor{\kappa}, \tensor{\sigma})$, as well as the instantaneous (test) strain-rate and stress tensors at the position of particle $i$, $\test{\vec{x}} = (\test{\tensor{\kappa}}, \test{\tensor{\sigma}}) = (\tensor{\kappa}_i(t), \tensor{\sigma}_i(t))$.

We refer to this method, which uses a constitutive relation learned through a Gaussian Process regression scheme, within a macroscopic flow simulation, as a GP accelerated Multi-Scale Simulation, or GP-MSS. We are mainly interested in learning this constitutive relation from microscopic polymer simulations, but to check the consistency of this approach, we will also consider learning from the constitutive relations themselves (e.g., from the upper-convected Maxwell model).

\subsection{Algorithmic Complexities}
To understand the benefits of our proposed GP-MSS approach with respect to a standard MSS, we should consider the complexities of both algorithms. Since both methods employ the same macroscopic description for the flow, any gains or losses will be found in the calculation of the stresses. For the MSS, this is done by solving for the microscopic dynamics of the polymer chains. Assuming a coarse-grained entanglement model, the time and memory requirements will scale as $\scaling{N_f \times N_p \times z}$, with $z$ the number of entanglement points per chain (for the case of non-interacting dumbbells we have $z=1$). Our experience shows that $N_p\times z$ should be of the order of $10^{4}-10^{5}$ or larger, in order to obtain reliable stress measurements. The complexity associated to the GP learning procedure is divided in two: the training step and the prediction step. The former is the most expensive of the two, scaling in time as $\scaling{n_{\text{training}}^2}$, and in memory as $\scaling{n_{\text{training}}}$\cite{Gardner2018,Wang2019}. However, we note that this procedure only needs to be done once, the resulting constitutive relation can then be used to perform arbitrarily complex flow simulations. To evaluate the performance of the GP-MSS, we focus then on the cost of making new predictions to update the stresses, as given by Eq.~\ref{e:gpmss_sigma}. This involves evaluating the average of the conditional GP at the test positions $\test{\vec{X}}$, corresponding to the values of $\tensor{\kappa}$ and $\tensor{\sigma}$ for each fluid particle. From Eq.~\eqref{e:gp_cond}, we see that we must compute expressions of the form $K(\test{\vec{X}}, \vec{X})\cdot \left(\inv{K(\vec{X}, \vec{X})} \cdot \delta\vec{f}(\vec{X})\right)$, where the term in parenthesis depends only on the training points $\vec{X}$. This term, which evaluates to a vector of length $n_{\text{training}}$, can also be precomputed. The prediction step can then be reduced to a simple matrix-vector multiplication, which scales as $\scaling{N_f\times n_{\text{training}}}$. However, the memory requirements are still only $\scaling{n_{\text{training}}}$.

The asymptotic ratio of the GP-MSS to MSS stress calculation time is then $\scaling{n_{\text{training}}/(N_p \times z)}$, which highlights the importance of generating a minimal high-quality training data set. The rule-of-thumb is to have the number of training points be smaller than the number of degrees of freedom in the microscopic polymer chain simulator. For the MSS simulations considered here, we need $N_p \simeq 10^{5}$ dumbbells to obtain reliable flow predictions. Using a random sampling protocol to generate the training data, we can achieve this same level of accuracy using $n_{\text{training}}\simeq 10^{3}$ training point. Although this does not account for the constant proportionality factors associated to each of the calculation costs, this two order of magnitude speedup is in agreement with the results of our simulations. 

The issue of generating an optimal training dataset, which in this case will maximize the time-saving with respect to the MSS, while keeping the same level of accuracy, is at the heart of most ML problems. 
It is quite likely such a protocol will have to be tailored to the specific microscopic model one wishes to study, as well as the macroscopic flow regimes to be simulated. For the systems studied here, random sampling proved to be enough, but it is not sure whether or not this will generalize to more complex setups.

\section{Results}\label{s:results}
\begin{figure}[ht!]
  \includegraphics[width=\columnwidth]{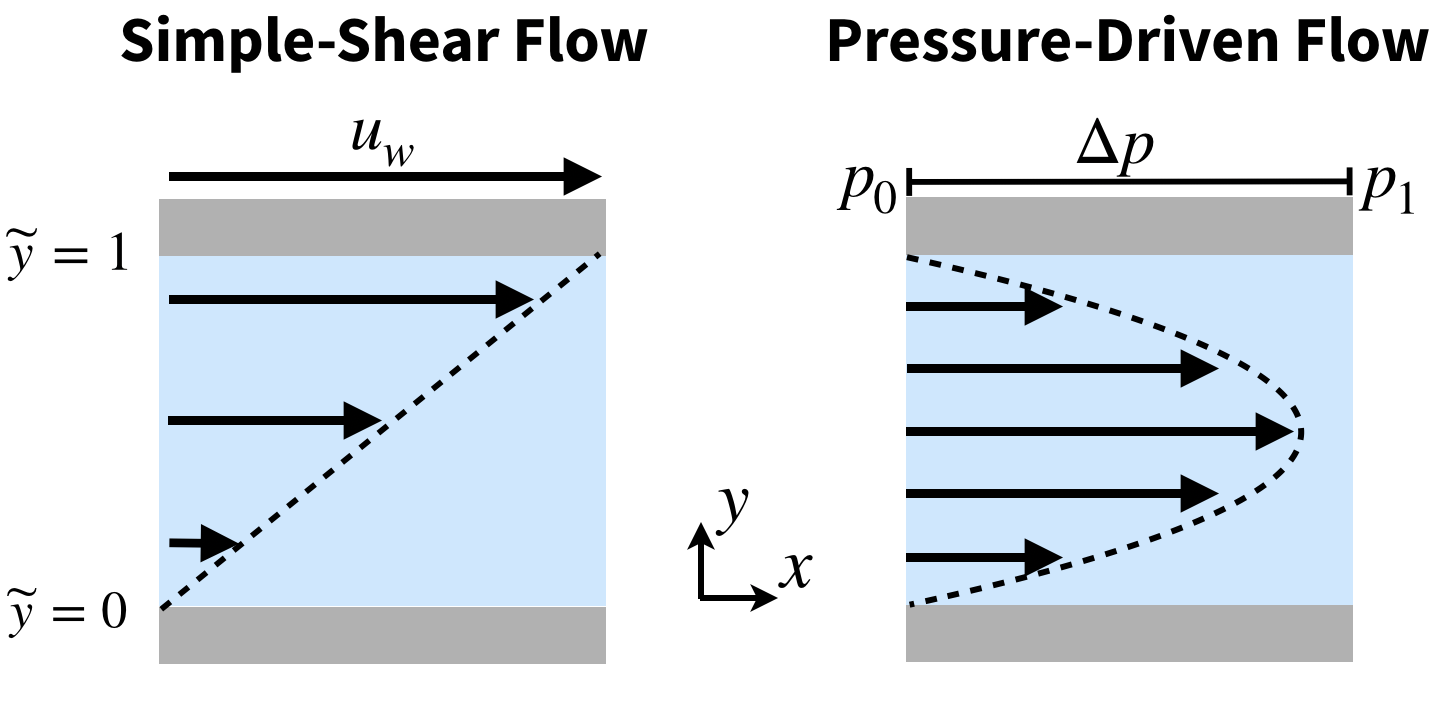}
  \caption{\label{f:flows}Schematic representation of the two standard flow problems we have used to test our learning strategy, simple-shear flow and pressure driven flow. For the former, we consider a simplified 1D description, while for the latter we take into account the full 2D nature of the stress and strain-rate tensors.}
\end{figure}

\begin{figure}[ht!]
  \includegraphics[width=0.9\columnwidth]{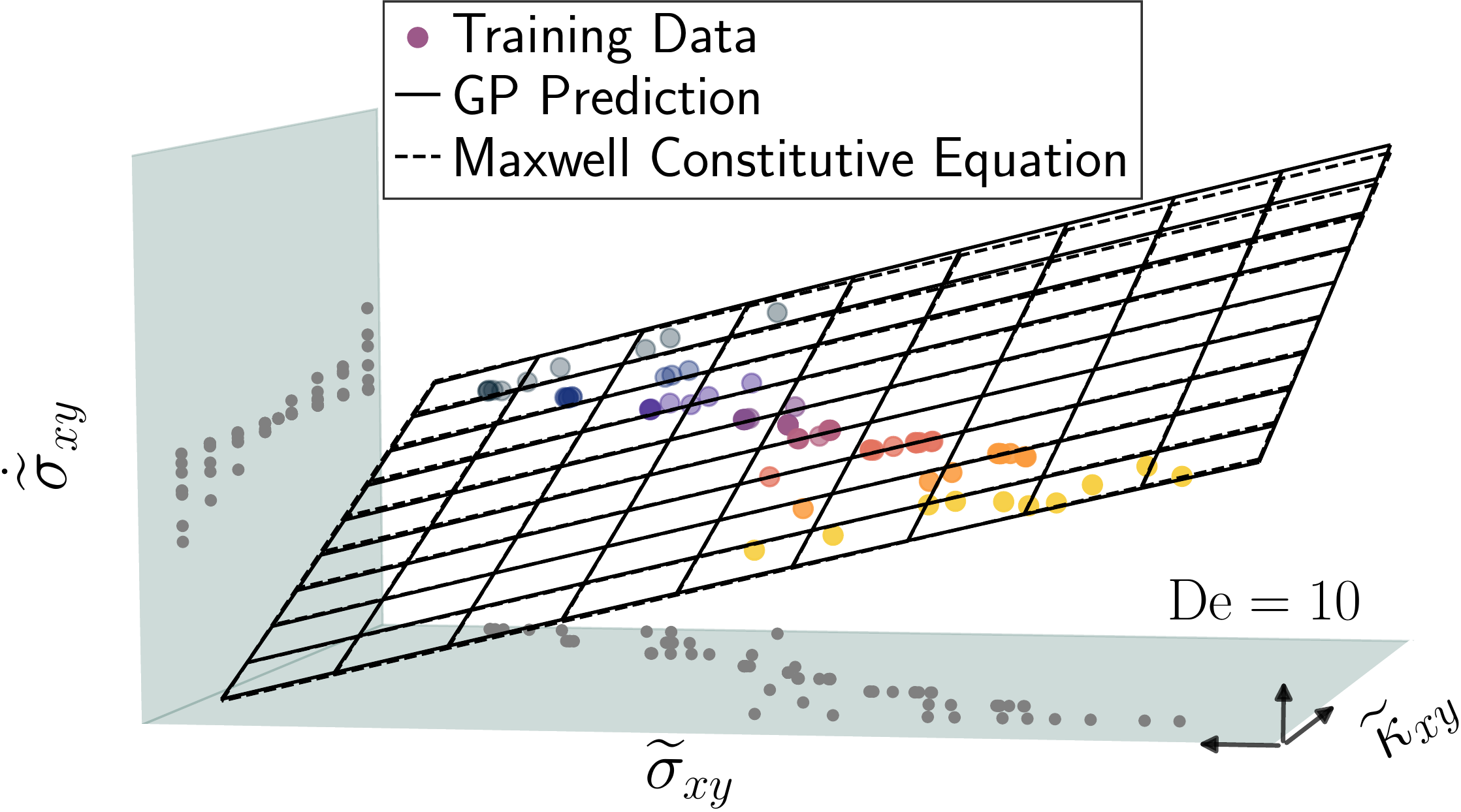}\\
  \includegraphics[width=\columnwidth]{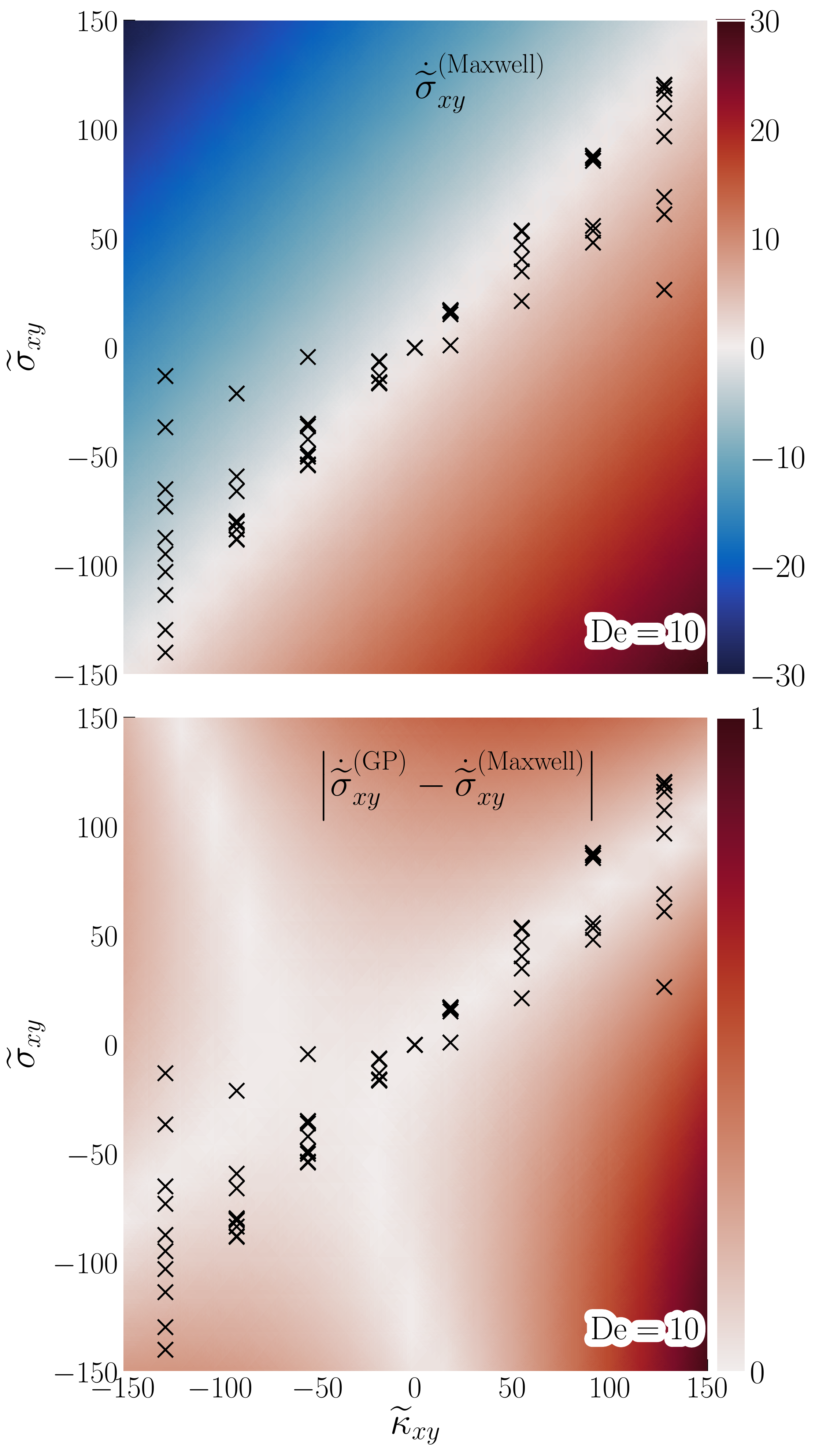}
  \caption{\label{f:1d_learn}(color online) The learned constitutive relation for the 1D simple shear flow problem at $\De = 10$ using $N_p = 10^3$ dumbbells for the microscopic model. (top) Training points, GP prediction, and the exact Maxwell constitutive equation ($N_p = \infty$ limit). (middle) Color map showing the exact constitutive relation $\dot{\adim{\sigma}}_{xy}^{\text{Maxwell}}$, as a function of $\adim{\kappa}_{xy}$ and $\adim{\sigma}_{xy}$. (bottom) Color map showing the absolute error between the GP prediction and the exact solution. The markers in the bottom two graphs show the location of the training data set.}
\end{figure}

We will consider two basic flow problems in order to validate our proposed learning strategy (see Fig.~\ref{f:flows}): (1) simple-shear flow and (2) pressure driven flow. Given their symmetry, the flow in both systems is effectively one-dimensional, but a complete description of their dynamics requires that we account for all components of the stress, and their coupling to the flow. Furthermore, if this approach is to be applied to general geometries, it should be capable of learning the appropriate form of the constitutive relation for the stress without any simplifying assumptions (although it can be useful to introduce this additional information in specific cases). To show how our learning procedure can be extended as the dimensionality of the system increases, we will study (1) the simple-shear flow problem in 1D and the (2) pressure-driven flow problem in 2D. 

For the simple-shear flow case we assume that $v_y = 0$ and the strain-rate and stress tensors have only one non-zero component, i.e., the $xy$ component. The learning problem is then 2D, since the constitutive relation is of the form $\dot{\adim{\sigma}}_{xy}(\adim{\kappa}_{xy}, \adim{\sigma}_{xy})$. For the pressure-driven case we consider the full 2D nature of the stress and strain-rate dependence, such that the learning problem is now $7$-dimensional, $\dot{\adim{\sigma}}_{\alpha\beta}(\adim{\kappa}_{xx}, \adim{\kappa}_{xy}, \adim{\kappa}_{yx}, \adim{\kappa}_{yy}, \adim{\sigma}_{xx}, \adim{\sigma}_{xy}, \adim{\sigma}_{yy})$.
For the flows we are considering $\adim{\kappa}_{\alpha z} = \adim{\kappa}_{z \alpha} = 0$, and since $\trace{(\adim{\tensor{\kappa}}) = 0}$, we have that $\adim{\kappa}_{xx} = -\adim{\kappa}_{yy}$. Instead of explicitly introducing such relationships into the model, we have preferred to learn them directly from the training data.
As we have chosen a system of non-interacting Hookean dumbbells for our microscopic polymer model, we are able to compare our results with the exact analytical constitutive equation, Eq.\eqref{e:maxwell}. Thus, we can check the convergences and sensitivity of our results, both in terms of the number of dumbbells $N_p$ used in the microscopic simulations, as well as the number of training points used in the GP regression $n_{\text{training}}$. Finally, since the absolute value of the local stress-tensor is not a physically measurable quantity, as opposed to the forces or velocities, we prefer to evaluate the accuracy of the learning strategy by comparing the error in the macroscopic predictions for the forces in the fluid and the flow velocities.

\subsection{Simple-Shear Flow (2D learning)}\label{s:results_1d}
To arrive at an effective 1D description for this simple-shear flow problem we have assumed the stress is a function of $y$ only, and that the system starts from a relaxed state $\adim{\sigma}_{\alpha\beta}(t= 0)= 0$. The Maxwell constitutive relation, Eq.~\eqref{e:maxwell}, then takes the following form
\begin{align}
  \dot{\adim{\sigma}}_{xy}(\adim{y}, \adim{t}\,) & = \frac{1}{\De}\bigg[ \adim{\kappa}_{xy}\big(\adim{y},\adim{t}\,\big) - \adim{\sigma}_{xy}\big(\adim{y}, \adim{t}\,\big)\bigg]\label{e:sigmadot_maxwell1d}
\end{align}
where $\adim{\sigma}_{xx}(t) = \adim{\sigma}_{yy}(t) = \adim{\sigma}_{zz}(t) = \adim{\sigma}_{xz}(t) = \adim{\sigma}_{yz}(t) = 0$.

Microscopic training data was generated by performing fixed shear-rate simulations for a system of $N_p=10^3, 10^4, 10^5$ non-interacting dumbbells in 3D for two different Deborah numbers, $\De = 1$ and $10$. In each case, we used nine different values of the adimensionalized shear-rate $\adim{\kappa}_{xy}$ within the range $[-150, 150]$, including $\adim{\kappa}_{xy} = 0$. The time step was set to $\Delta\adim{t} = 10^{-4}$ and the simulations were performed up to a maximum time of $\subt{\adim{t}}{max} = 5 \De$. The trajectory of the stress $\adim{\sigma}_{xy}(t)$ was then smoothed using a Gaussian filter with a width of $5\times 10^{-3} \tau^{(m)}$, in order to reduce the noise in the estimates for $\dot{\adim{\sigma}}_{xy}$. The time-derivative of the stress, obtained from the smoothed data, was computed using the following finite-difference approximation
\begin{align}
  \dot{\adim{\sigma}}_{xy}(\adim{y}, \adim{t}) & = \frac{\adim{\sigma}_{xy}(\adim{t}) - \adim{\sigma}_{xy}(\adim{t}-\Delta \adim{t})}{\Delta\adim{t}}
\end{align}
The resulting $(\adim{t}, \adim{\sigma}_{xy}, \dot{\adim{\sigma}}_{xy})$ data is then partitioned into ten (possibly overlapping) randomly selected intervals over $\adim{\sigma}_{xy}$, of width $(\max{(\adim{\sigma}_{xy})}-\min{(\adim{\sigma}_{xy})})/10$. Averaging over the points contained in each interval allows us to define a corresponding pair of input $\vec{x} = (\adim{\kappa}_{xy}, \adim{\sigma}_{xy})$ and output $f(\vec{x}) = \dot{\adim{\sigma}}_{xy}$ training points. The variance $\epsilon^2$ in $\dot{\adim{\sigma}}_{xy}$, within each interval, is used as an estimate of the measurement error, and added to the corresponding covariance sub-matrix $K(\vec{X},\vec{X})$, in order to perform the GP regression (see Eq.~\ref{e:gp_joint}), i.e., $K(\vec{X},\vec{X})_{ij} = k(\vec{X}_i, \vec{X}_j) + \epsilon_i^2\delta_{ij}$.

We use a product kernel for this 2D regression problem $\vec{x} = (x_1, x_2) = (\adim{\sigma}_{xy}, \adim{\kappa}_{xy})$, such that
\begin{align}
  k(\vec{x}, \vec{x}^\prime) & = \Gamma^2 k_{\text{SE}}(x_1, x_1^\prime; l_1) \cdot k_{\text{SE}}(x_2, x_2^\prime; l_2)\label{e:gp_k1d}
\end{align}
We thus have three hyper-parameters, $\Gamma$, and the two length scales for $\adim{\sigma}_{xy}$ and $\adim{\kappa}_{xy}$, so that $\vec{\Theta} = (\Gamma, l_1, l_2)$. We use independent and uniform logarithmic priors, $p(\ln{\Theta}_i)\propto \text{const}$, which are scale-invariant, such that
\begin{align*}
  p(\vec{\Theta}) & = p(\Gamma) \cdot p(l_1)\cdot p(l_2)\propto \frac{1}{\Gamma}\cdot\frac{1}{l_1}\cdot\frac{1}{l_2}
\end{align*}
To train the model, we performed Hamiltonian Monte-Carlo simulations, using the No-U-Turn Sampler (NUTS), with the PyMC3 python package\cite{Betancourt2017, Hoffman2014, Salvatier2016}. The optimal parameters $\bm{\Theta}_0$ are then taken to be the averages over the posterior distribution $\vec{\Theta}^{\text{avg}}$, and used to define the conditional GP for the constitutive relation. Fig.~\ref{f:1d_learn} shows the results of this learning procedure for the case of $\De = 10$. We are able to learn the Maxwell constitutive relation, Eq.~\eqref{e:sigmadot_maxwell1d}, providing excellent predictions over a wide range of parameters. However, the prediction error increases the farther we get from the training points, as expected. This highlights the importance of generating a well chosen training data set, representative of the region in function space one is interested in. It is reassuring to note that even with the naive random sampling strategy outlined above, we are able to obtain precise predictions for the constitutive relation. While this is due to the simplicity of the function, we will show below that this approach also generalizes to more complicated functional forms and higher dimensions.
\begin{figure*}[hb!]
  \includegraphics[width=0.95\textwidth]{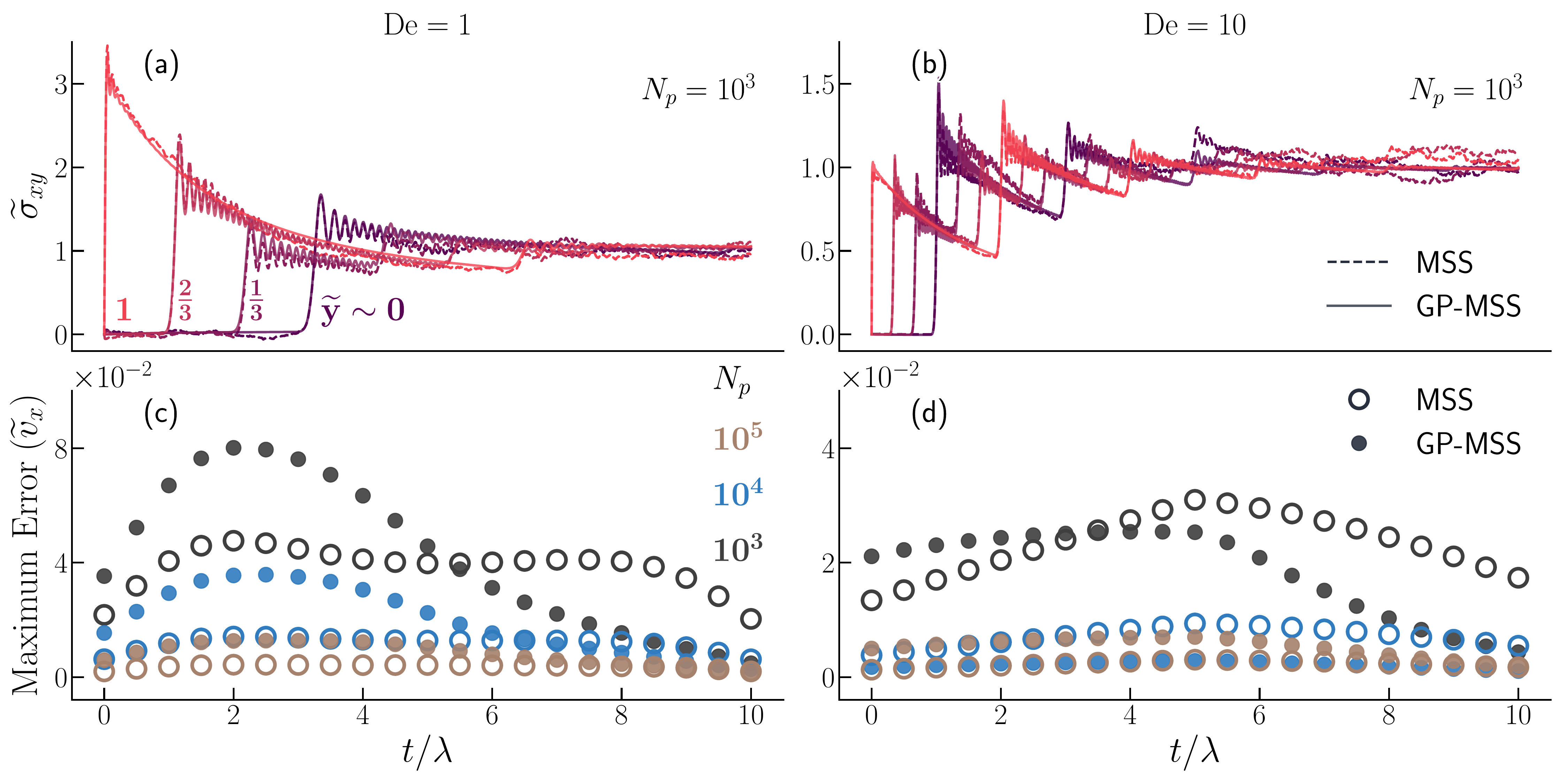}
  \caption{\label{f:1d_sigma+vel} (color online) (a-b) Time evolution of the stress obtained from simple-shear flow simulations ($\Rey=10$) with the learned constitutive equation (GP-MSS) as well as those from the full MSS, using microscopic dumbbell simulators ($N_p = 10^{3}$). Dark (light) colored lines correspond to small (large) values of the channel height $\adim{y}$. At $t=0$, a velocity wave starts at $\adim{y}=1$ and propagates down the channel, bouncing back at the lower wall.(c-d) Maximum absolute error in the velocity, for all points in the channel, obtained from the MSS and GP-MSS predictions. The error is computed with respect to the results of the (exact) Maxwell constitutive equations, $\abs{\adim{v}_x - \adim{v}_x^{\text{(Maxwell)}}}$. Results for different number of dumbbells $N_p = 10^3, 10^4, 10^5$ used in the microscopic simulations are shown.}
\end{figure*}
Using this learned constitutive relation, we performed flow simulations under simple shear, at $\Rey = 10$, and compared the results to those obtained from standard MSS (with microscopic dumbbell simulators), as well from the Maxwell constitutive equation.

Considering the one-dimensional nature of the flow problem, one can use a simple Eulerian description instead of the Lagrangian one. Thus, the system was discretized in the vertical $y$ direction, using $128$ grid points, with the velocity of the top and bottom walls set to $\adim{v}=1$ and $\adim{u}=0$, respectively. Starting from a quiescent fluid, a velocity wave will start at the top wall, and propagate through the channel, bouncing back and forth at the walls, before a steady-state linear velocity profile is obtained\footnote{\label{Note1}See Supplemental Material SM1~and~SM2 at [URL] for the time-evolution of the velocity profile, for the simple-shear flow case, obtained from MSS and GP-MSS, as well as the exact solution given by the Maxwell constitutive relation. Results for $\De=1$ (SM1) and $\De=10$ (SM2) are provided}. This transient regime is more pronounced at high \De, as evidenced by the three kicks in the time-evolution of the stress at $\De=10$, corresponding to the arrival of the wave-front. For comparison, at $\De=1$ only one kick is observed (see Fig.~\ref{f:1d_sigma+vel}~(a-b)). The excellent agreement obtained between the MSS and GP-MSS predictions for the stress is further evidence of the success in learning the constitutive relation. The stress fluctuations obtained from the MSS at long times are a consequence of the large statistical fluctuations that come from using a finite number of dumbbells. It is encouraging to see that the GP-MSS predictions do not show such behavior. This is because our learning strategy properly accounts for the measurement error in the training data, allowing us to infer the ``true'' function. To further test our ability to capture the history-dependence of the flow, we have also considered an oscillatory shear flow (not shown), and obtained a similar level of agreement between MSS and GP-MSS predictions. We note that a generalized Newtonian approach, which assumes the stress $\tensor{\sigma}$ is a function of $\tensor{\kappa}$, would not be able to capture this memory effect\footnote{See Supplemental material SM3 at [URL] for the time-evolution of the velocity profile, for oscillatory-shear flow, obtained from MSS and GP-MSS, as well as the exact solution given by the Maxwell constitutive relation. Simulations are performed by setting the velocity of the top wall to be $v_x = U\cos{(\omega t)}$, where the magnitude $U$ and frequency $\omega$ of the shear flow are set by the Deborah and (squared) Womersley numbers, $\textrm{De} = 1$ and $\textrm{Wo}^2=L^2\rho\omega/\eta=20$, respectively. For comparison purposes, we have also shown simulation results for a corresponding generalized Newtonian fluid, i.e., assuming a constitutive relation of the form $\sigma_{xy} = \eta^{\text{(eff)}}(\dot{\gamma}) \cdot \dot{\gamma}$ (with $\dot{\gamma}$ the shear rate, and $\eta^{\text{(eff)}}$ the effective viscosity), for which there is no memory effect}.

Fig.~\ref{f:1d_sigma+vel}~(c-d) shows the maximum absolute error in the predicted velocities, among all points in the system, as a function of the number of dumbbells $N_p$ used in the microscopic simulations. This error is evaluated with respect to the velocities obtained from macroscopic simulations with the exact constitutive relation. Not surprisingly, given the good agreement in the stress profiles, the velocities also coincide. There is however a small offset in the transient regime\cite{Note1}, with the GP-MSS velocity wave showing a slight delay (advance) with respect to the MSS or Maxwell predictions at $\De = 1$ ($10$). This is the main source error reported in Fig.~\ref{f:1d_sigma+vel}~(c-d), and is due to the small number of training data around this particular point in $(\adim{\tensor{\kappa}},\adim{\tensor{\sigma}})$ space. Two aspects deserve to be highlighted: First, in all cases we have considered, the error of the GP-MSS results is of the same order of magnitude as the MSS results. While the error can be two times larger, for times near the beginning of the start-up flow, it can also be considerably smaller, particularly at steady-state. This is most obvious for the case with small number of dumbbells $N_p = 10^3$. Second, the error decreases considerably as $N_p \rightarrow \infty$, as this decreases the statistical errors in both the MSS and the training data used to learn the constitutive relations. However, we note that the accuracy of the GP-MSS will also be affected by the quality of the training points. In the case of $\De=10$, for example, we happened to obtain slightly better results for $N_p = 10^4$ than for $N_p = 10^5$.

\subsection{Pressure-Driven Flow (7D learning)}\label{s:results_2d}\
\begin{figure}[ht!]
  \includegraphics[width=\columnwidth]{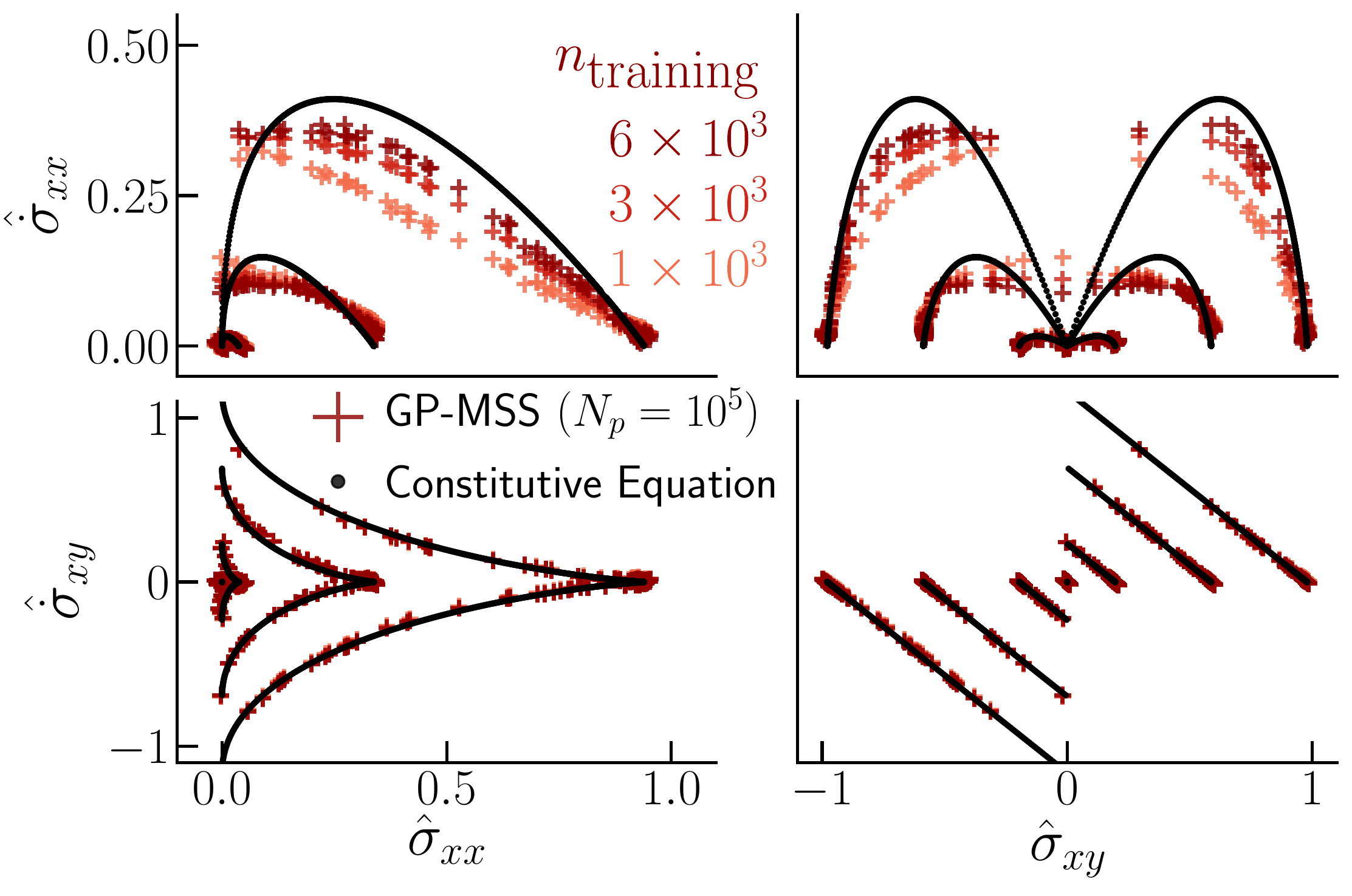}
  \caption{\label{f:2d_training}(color online) GP predictions (cross symbols) for the constitutive relation $\dot{\tensor{\sigma}}$, learned using $n_{\text{training}} = 1,3$~and~$6\times 10^{3}$ training points (generated from microscopic simulations with $N_p = 10^5$ dumbbells). Exact results (dot symbols) were obtained from the Maxwell constitutive relation, corresponding to the $N_p\rightarrow\infty$ limit. The data clearly shows that the results improve as the number of training points increases, as expected. A caret $\hat{\left(\cdot\right)}$ indicates data that has been scaled to lie in the range $[-1,1]$, with the input ($\kappa_{\gamma\delta}, \sigma_{\alpha\beta}$) and output ($\dot{\sigma}_{\alpha\beta}$) scaled separately, to facilitate visualization.}
\end{figure}

For the pressure driven-flow problem we will consider the full 2D nature of the system. The learning problem then consists of three GP regressions, one each for $\dot{\adim{\sigma}}_{xx}$, $\dot{\adim{\sigma}}_{xy}$, and $\dot{\adim{\sigma}}_{yy}$, all of them functions in a seven-dimensional space $\vec{x} = (\adim{\kappa}_{xx}, \adim{\kappa}_{xy}, \adim{\kappa}_{yx}, \adim{\kappa}_{yy}, \adim{\sigma}_{xx}, \adim{\sigma}_{xy}, \adim{\sigma}_{yy})$. Given the increased complexity with respect to the effective 1D flow considered previously, which resulted in a 2D learning problem, we have simplified the procedure to generate the training data. This is not directly related to the type of flow, but rather to the dimensionality of the problem. Microscopic 3D simulations for $N_p\lesssim 10^5$ non-interacting dumbbells at fixed strain-rate are performed. We considered both simple-shear and planar elongational flows within the range $\abs{\adim{\kappa}_{\alpha\beta}}\le 150$. We chose $11$ different values for each of $\adim{\kappa}_{xx}$, $\adim{\kappa}_{xy}$, and $\adim{\kappa}_{yy}$, in addition, results for $\adim{\kappa}_{yx}$ were obtained by taking the transpose of those at $\adim{\kappa}_{xy}$.  The simulation time step was fixed to $\Delta \adim{t} = 10^{-6}$, the maximum time was set to be $\subt{\adim{t}}{max} = 10\De$, and the corresponding $\tensor{\sigma}(t)$ trajectories where saved and used to generate the training data. For this, we randomly selected $N\simeq 10^{3}$ points $(\adim{t},\adim{\kappa}_{\gamma\delta}, \adim{\sigma}_{\alpha\beta})$, and included the initial state $\adim{\sigma}_{\alpha\beta}(t=0) = 0$. Then, we use each point to define a time interval of width $0.1\micro{\tau}$, with $\micro{\tau}$ the polymer relaxation time. For the training data, we take the stress to be the average stress within the time interval, whereas the time derivative of the stress $\dot{\adim{\sigma}}_{\alpha\beta}$ is taken to be the difference at the two end-points, i.e., using a coarse-grained time-step of $\Delta \supt{t}{(c)} \simeq 0.05 \De$. The measurement error for $\dot{\adim{\sigma}}_{\alpha\beta}$ cannot be ignored, but it is also not easy to evaluate in this high-dimensional space. Therefore, we introduce this error $\epsilon_{\alpha\beta}$ as an additional hyper-parameter that should be learned from the data. It is in such situations where the benefit of adopting a Bayesian approach pays off.

\begin{figure}[ht!]
  \includegraphics[width=\columnwidth]{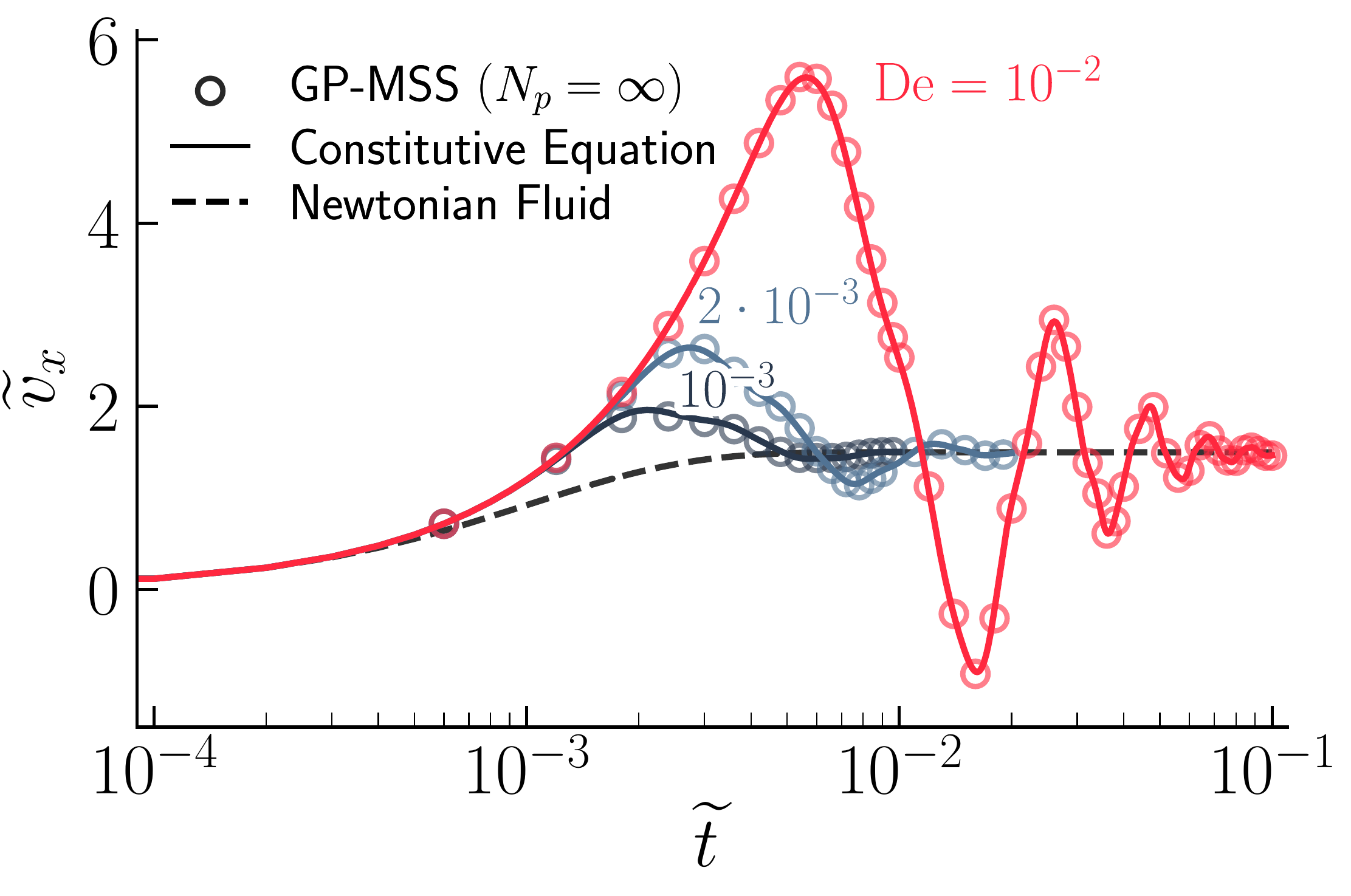}
  \caption{\label{f:2d_vde}(color online) Time evolution of the velocity, at $\adim{y} = 0.5$, for the pressure-driven flow problem ($\Rey=10^{-2}$). Results for three different $\De$ are shown, together with the corresponding Newtonian fluid ($\De = 0$). Results obtained using the (solid line) exact Maxwell constitutive equation are compared with (open symbols) GP-MSS using a constitutive relation learned from ($n_{\text{training}} = 10^{3}$) noiseless training points generated from the Maxwell model.}
\end{figure}
To summarize, we place a GP prior on each $\dot{\sigma}_{\alpha\beta}$, such that $\dot{\sigma}_{\alpha\beta} = f(\vec{x}) \sim \gp(0, K(\vec{X},\vec{X}))$. For the covariance matrix, we have $K(\vec{X}, \vec{X})_{ij} = k(\vec{X}_i, \vec{X}_j) + \epsilon^2 \delta_{ij}$, with the measurement error assumed to be constant, equal for all $\vec{x}$. Here, we have used an additive first-order kernel, such that
\begin{align}
  k(\vec{x}, \primed{\vec{x}}) & = \Gamma^2\sum_{\Lambda} k^{(\Lambda)}(x_\Lambda, \primed{x}_\Lambda; l_\Lambda)
\end{align}
with $\Lambda = 1,\cdots,7$ specifying one of the seven possible input components of $\vec{x} = (\adim{\kappa}_{xx}, \adim{\kappa}_{xy}, \adim{\kappa}_{yx}, \adim{\kappa}_{yy}, \adim{\sigma}_{xx}, \adim{\sigma}_{xy}, \adim{\sigma}_{yy})$. The choice of an additive kernel is motivated by the fact that it allows for non-local interactions, making it less susceptible to the \textit{curse of dimensionality} and allowing for better extrapolation in regions far way from the training data. In contrast, a multiplicative Kernel will rapidly revert to the mean away from the training data\cite[{Ch.2.4 and 6.1}]{Duvenaud2014}. This is not an issue for the 2D learning problem considered previously, as it was easy to generate a large enough sample of training points. For each 1D kernel $k^{(\Lambda)}$ we have one associated hyper-parameter or length-scale $l_\Lambda$, for a total of seven length-scale hyper-parameters. Together with $\Gamma$ and the (unknown) measurement error $\epsilon_{\alpha\beta}$ associated to the training data for $\dot{\adim{\sigma}}_{\alpha\beta}$, this results in a total of nine hyper-parameters needed to learn each $\dot{\adim{\sigma}}_{\alpha\beta}$. Note that we are assuming that the measurement error is constant, i.e., it does not depend on $\vec{x}$, although it can be different for the different components of the constitutive relation. The optimal values were obtained by maximizing the log-posterior, Eq.~\eqref{e:logp_theta}, assuming a constant prior for the hyper-parameters ($p(\vec{\Theta}) = \text{const}$). For this, we use ADAM\footnote{We mainly used the default parameters proposed by Kingma and Ma for Machine Learning problems, namely, the step size is $\alpha=10^{-2}$, the hyper-parameters are $\beta_1 = 0.9$, $\beta_2 = 0.999$, and $\varepsilon = 10^{-8}$. We have set the maximum number of iterations to be $500$}, a stochastic gradient descent algorithm\cite{Carleo2019,Kingma2015}, as implemented in GPyTorch\cite{Gardner2018,Wang2019}.

This learning procedure results in three distinct functions $\dot{\adim{\sigma}}_{xx}$, $\dot{\adim{\sigma}}_{xy}$, and $\dot{\adim{\sigma}}_{yy}$ of seven variables. The most interesting, non-trivial components of the constitutive relations (for the flows we are considering) are the $xx$ and $xy$ components, visualized in Fig.~\ref{f:2d_training} for three different number of training points $n_{\text{training}} = 1,3$~and~$6\times 10^{3}$ generated from microscopic simulations of $N_p = 10^{5}$ dumbbells\footnote{See Supplemental Material SM4 at [URL] for the full constitutive relation map used in the learning procedure for the 2D pressure driven flow problem, for the case of $n_{\text{training}} = 1\times 10^{3}$ points, together with data generated from the exact Maxwell constitutive relation. In addition, we also plot the trajectory data $(\tensor{\kappa}(t), \tensor{\sigma}(t), \dot{\tensor{\sigma}}(t))$ obtained from GP-MSS ($\De = 1\times 10^{-2}$) for three representative points $\tilde{y}\sim 0, 1/4, 1/2$ along the channel. Simulation results using the exact constitutive relation are also given}. As for the 1D problem considered above, we compare our results with data generated from the exact Maxwell constitutive relation ($N_p \rightarrow \infty$), and, as expected, obtain better agreement as $n_{\text{training}}$ increases. Finally, for comparison purposes, we have also learned the constitutive relation from training data generated from the exact solution, i.e., the Maxwell model ($N_p\rightarrow\infty$).

\begin{table}[ht!]
  \begin{tabular}{lll}
    \hline
    Parameter & Description & Value\\
    \hline    
    \Rey & Reynolds number &$10^{-2}$ \\
    \De  & Deborah number  &$10^{-3} - 10^{-1}$ \\
    \Cs  & artificial sound-speed & $10$ \\
    $\Delta\adim{t}$ & time-step & $10^{-6}$ \\
    $N_p$ & Number of dumbbells (MSS) & $10^{3}-10^{5}$ \\    
    $N_f$& No. of fluid particles & $540 = 18\times30$ \\
    $(\adim{L}_x,\adim{L}_y)$ & System Size & $(0.6, 1.0)$ \\    
    $\adim{b}$ & Initial particle size & $1/30$\\
    $\adim{h}$ & Smoothing length & $ 1.5\, \adim{b}$\\
    $\adim{R}_c$ & Cut-off length & $3\, \adim{h}$\\
    $\adim{\vec{F}}_e$ & External driving force & $(12, 0)$\\
    $n_{\text{training}}$ & GP training points & $1\times 10^{3} - 6\times 10^{3}$\\
    $\Delta \adim{t}^{\,\text{(c)}}$ & GP coarse-graining window  & $0.05\De$
  \end{tabular}
  \caption{\label{t:2d_params}Default (non-dimensionalized) parameter values used for the $2$D MSS and GP-MSS simulations.}
\end{table}
We used the learned constitutive relations to perform 2D simulations for the pressure driven flow problem, comparing our GP-MSS results with those of the conventional MSS and the exact Maxwell constitutive relation. We perform SPH simulations for all three cases, in a channel whose width $\adim{L}_x=0.6$ (along the flow direction) is about half its height $\adim{L}_y=1$. The fluid is discretized using $540$ particles, initially arranged on a regular lattice within the system, corresponding to a $18\times 30$ (width $\times$ height) array of particles. The initial distance between each particle is $\adim{b}=1/30$, the smoothing length is $\adim{h}=1.5 \adim{b}$, and the cut-off length is $\adim{R}_c = 3\adim{h}$. We performed simulations at $\Rey=10^{-2}$, for $\De \lesssim 10^{-2}$, which is enough to observe elastic effects in the flow. Fig.\ref{f:2d_vde} shows the time-evolution of the velocity at the center of the channel ($\adim{y}= 0.5$) for $\De = 1, 2 $~and~$10\times 10^{-3}$, as obtained from GP-MSS using the constitutive relation learned from exact training data ($N_p\rightarrow\infty$), SPH simulations using the Maxwell constitutive equation, and a corresponding Newtonian fluid ($\De = 0$). The default parameters for these $2D$ (SPH) MSS and GP-MSS are summarized in Table~\ref{t:2d_params}.

As expected, at steady-state the system is indistinguishable from a Newtonian fluid. However, at short times $\adim{t}\lesssim 0.1$ effects due to the elastic energy stored in the dumbbells are visible. This is seen in the velocity oscillations around the Newtonian values, for which the overshoot can result in speeds that are more than three times the steady-state value. These elastic effects become more important as $\De$ is increased, and even results in large negative (transient) velocities at $\De \gtrsim 10^{-2}$. The differences between the GP-MSS results, using the learned constitutive equation, and those from the exact constitutive equation are negligible. While these GP-MSS relied on a constitutive relation learned from exact noiseless data ($N_p\rightarrow\infty$), using a finite number of dumbbells $N_p$ to generate the training data yields similar level of agreement, as will be shown below.

\begin{figure}[ht!]
  \includegraphics[width=0.8\columnwidth]{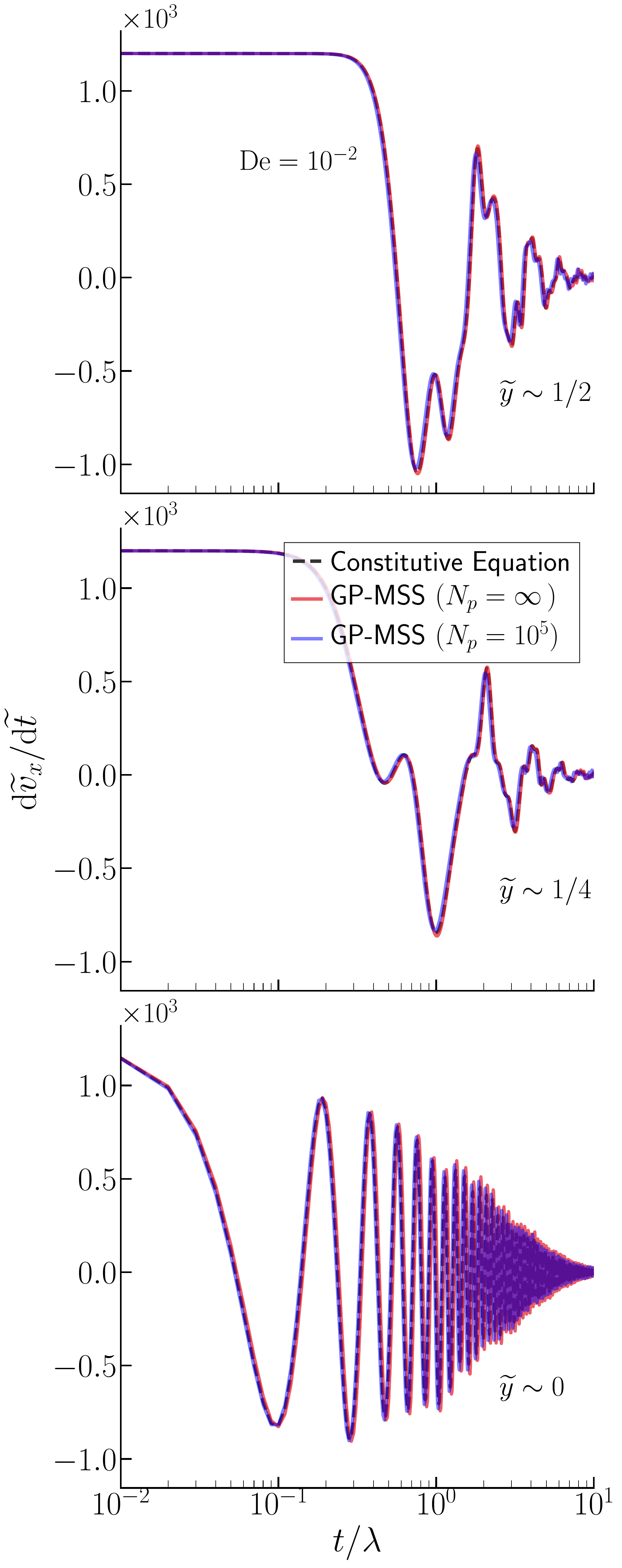}
  \caption{\label{f:2d_dvdt_De-2}(color online) Scaled force at three locations along the height of the channel, at $\adim{y}\sim 1/2$, $1/4$, and $0$, for the pressure-driven flow problem ($\Rey=10^{-2}$). Results are from GP-MSS, using a constitutive relation obtained from microscopic simulations ($N_p=10^5$ dumbbells), as well as from the exact constitutive equation ($N_p\rightarrow\infty$), with $n_{\text{training}} = 10^{3}$. Exact results, obtained from simulations using the Maxwell constitutive relation are also shown.}
\end{figure}
\begin{figure}[ht!]
  \includegraphics[width=\columnwidth]{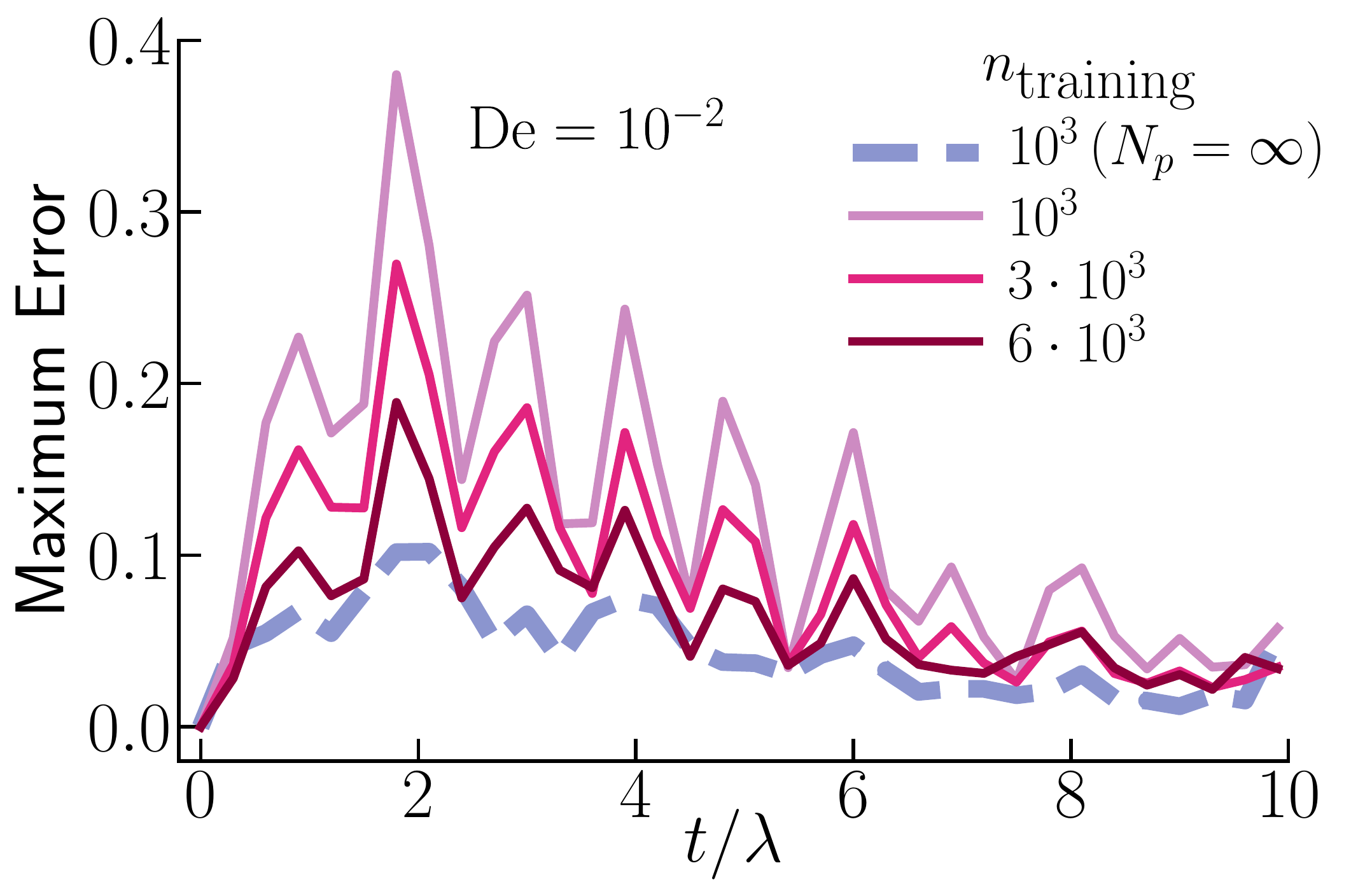}
  \caption{\label{f:2d_err}(color online) Maximum absolute error in the velocity obtained from GP-MSS of the pressure-driven flow, $\max{\left(\adim{v} - \adim{v}^{\text{(Maxwell)}}\right)}$, using constitutive relations learned on different number of training points (generated from microscopic simulations using $N_p=10^5$ dumbbells). Results using a constitutive relations learned from the exact solution ($N_p=\infty$) are also shown. The simulations were performed at $\Rey=\De=10^{-2}$.}
\end{figure}

First, although there is a small difference between the learned constitutive relation and the exact solution (see Fig.~\ref{f:2d_training}), particularly for the $\adim{\sigma}_{xx}$ component, the macroscopic predictions are in excellent agreement. This can be seen when looking at the forces in the fluid, as shown in Fig.~\ref{f:2d_dvdt_De-2} for three different positions along the height of the channel. Indeed, GP-MSS results using constitutive equations learned ($n_{\text{training}}=10^{3}$) from the exact solution ($N_p\rightarrow\infty$) or from microscopic simulations ($N_p= 10^5$) are indistinguishable from each other at this scale, and they coincide with macroscopic simulation results using the Maxwell constitutive relation, although there is a small lag in the forces for the $N_p=10^5$ MSS case. Second, we show that increasing the number of training points results in more accurate constitutive relations, and thus more reliable macroscopic flow simulations. We used the three constitutive relations of Fig.~\ref{f:2d_training}, generated from $n_{\text{training}} = 1,3$~and~$6\times 10^{3}$ training points, to perform GP-MSS, and compared the predicted velocity profiles with the exact solution, as given by SPH simulations using the Maxwell constitutive relation\footnote{See Supplemental Material SM5 at [URL] for the time-evolution of the velocity profile obtained from GP-MSS at $\De = 10^{-2}$ (open symbols), using constitutive relations learned from $n_{\text{training}} = 1,3,6\times 10^{3}$ points (generated from microscopic simulations with $N_p =10^{5}$ dumbbells), as well as the exact solution given by the Maxwell constitutive relation (solid line). Results obtained using a constitutive relation learned from $n_{\text{training}} = 1\times 10^{3}$ points generated from the exact constitutive relation are also shown (filled symbols)}. Fig.~\ref{f:2d_err} shows the maximum absolute error in the velocity as a function of time. Increasing the number of training points dramatically reduces the error in the simulations. Furthermore, the simulations using the constitutive relation learned on $N_p=10^{5}$ dumbbells give the same level accuracy as those using the constitutive relation learned from the exact solution. All things being equal, increasing the number of training points will give better results; however, what matters is the quality of the training data set. This is the reason why the best results are obtained with the constitutive relations learned from the exact data, even though only a relatively small number of training points are used.

\section{Conclusions}\label{s:conc}
We have developed a learning strategy that is able to infer the constitutive relation of polymer melt flows from a small number of microscopic or coarse-grained polymer simulations.
For this, we have used a Bayesian learning approach based on Gaussian Process (GP) regressions. GPs provide a probability distribution over functions, allowing us to infer the most likely values, given known training data. In addition, we can estimate the uncertainty in the predictions, as well as incorporate unknown or incomplete data (e.g., measurement errors). Previous work has shown how one can use this type of approach to learn the effective viscosity of a polymer melt flow, as a function of the local shear-rate\cite{Zhao2018}, or to parameterize a viscoelastic constitutive relation\cite{Zhao2020}. Here, we demonstrate that a learning scheme that includes memory effects can be developed, which is crucial in order to describe the flow dynamics of entangled polymers in complex flow geometries. This method has great potential for polymer processing, as it will allow us to consider flow problems relevant to industrial settings. In addition, we believe that similar learning strategies can be designed for other soft-matter systems, where the presence of multiple length- and time-scales gives rise to complex dynamical behavior that is expensive to simulate directly.

To validate the method, we have adopted the simplest possible microscopic polymer model, that of an ensemble of non-interacting Hookean dumbbells, since the exact constitutive equation is known in this case. This model was used in fixed strain-rate $\tensor{\kappa}$ simulations, under simple-shear and planar elongation, to generate the required training data. For this, the time-evolution of the stress $\tensor{\sigma}(t)$ was used to estimate the time-derivative $\dot{\tensor{\sigma}}$. Assuming that the constitutive relation can be written in differential form, as a function of the local strain-rate and stress, the goal is to learn the function $\dot{\tensor{\sigma}}(\tensor{\kappa}, \tensor{\sigma})$. Thus, the training data consists of input points $(\tensor{\kappa}, \tensor{\sigma})$ and the corresponding output $\dot{\tensor{\sigma}}$. We randomly selected a subset of $\simeq 10^{3}$ points and used them within a GP regression scheme in order to determine the optimal posterior distribution $p\pcond{\test{\dot{\tensor{\sigma}}}}{\tensor{\kappa}, \tensor{\sigma}, \test{\tensor{\kappa}}, \test{\tensor{\sigma}}}$ for the constitutive equation $\test{\dot{\tensor{\sigma}}}$ at new input test points $(\test{\tensor{\kappa}}, \test{\tensor{\sigma}})$, for which $\dot{\tensor{\sigma}}$ is not known. This conditional probability distribution is a GP, with average and covariance that are functions of both the training and test points.

We set the average $\avg{\test{\dot{\tensor{\sigma}}}}$, over the posterior distribution, to be our best prediction for the constitutive relation, and this function was then used within a macroscopic simulation in order to predict the flow behavior. In this way, we were able to carry out all our simulations at the macroscopic level, without having to impose any constitutive relation. Again, all we assumed was that the time-derivative of the stress is a function of the local stress and strain-rate tensors. The appropriate constitutive equation is learned from a (relatively) small number of microscopic simulations. The resultant method, which we have referred to as GP-MSS, gives results that are as accurate as conventional MSS, at a fraction of the cost. With our non-optimized python+numpy code, the difference in runtime between the full MSS (using $10^5$ dumbbells) and the GP-MSS (with $n_{\text{training}}=10^3$ training points) is $\sim 100$ times for the 2D pressure driven flow problem. We note that the GP-MSS run only marginally slower than simulations using the Maxwell constitutive relations. Thus, we get the best of both worlds, achieving run times comparable to macroscopic simulations, without sacrificing the accuracy provided by a microscopic polymer description. We expect the speedup afforded by the GP-MSS to improve dramatically when more realistic, and complex, microscopic polymer models are used. However, the reliability and efficiency of the GP-MSS will depend on the quality and size of the training dataset. Thus, care should be taken when devising the data generating protocol. As an added benefit to our approach, we note that it is also possible to maintain (consistent) information on the microscopic degrees of freedom, for example, by adopting a multi-fidelity representation\cite{Perdikaris2017,Raissi2017a}. In this case, a small number of microscopic simulators could be introduced in order to provide accurate (localized) stress measurements, which are then fused together with the approximate predictions provided by the learned constitutive relation. In future work we will apply this learning approach to tackle the problem of entangled polymer melt flows. This will allow us to consider 3D flows in complex geometries, which have so far remained out of reach for standard MSS techniques.
\appendix
\section{The SPH Method}\label{s:app_sph}
\subsection{Introduction}
The Smooth Particle Hydrodynamics (SPH) method, a particle-based method originally developed to solve astrophysics problems, provides a computationally convenient way to solve the Navier-Stokes equation in a Lagrangian framework\cite{Monaghan1992}. The fluid is discretized into a number $N_f$ of fluid particles that carry mass, momentum, and energy. Any function of the system can then be expressed as an interpolation over the (disordered) fluid particles, which serve as interpolation points. Consider a scalar quantity $A$ and a vector quantity $\vec{V}$, which depend on position. The value of $A\left(\vec{x}\right)$ or $\vec{V}(\vec{x})$, at a given position $\vec{x}$, which need not correspond to the position of any of the fluid particles, is given by
\begin{align}
  A\left(\vec{x}\right)   & = \int A\paren*{\vec{x}^\prime} W\paren*{\vec{x} - \primed{\vec{x}}, h} \df{\primed{\vec{x}}} \label{e:sph_a} \\
  \vec{V}\paren*{\vec{x}} & = \int \vec{V}\paren*{\primed{\vec{x}}} W(\vec{x}-\primed{\vec{x}}, h)\df{\primed{\vec{x}}}
\end{align}
with similar expressions for higher-order tensor fields. Here, $W(\vec{x},h)$ is a smoothing or interpolating kernel that should integrate to unity and tend to a delta function in the limit when the smoothing length $h$ goes to zero. In this work, we adopt a Gaussian interpolating kernel
\begin{align}
  W(\vec{r}, h)
  = \frac{1}{\paren*{h\sqrt{\pi}}^{D}}
  \exp\bracket*{-\frac{\norm{\vec{r}}^2}{h^2}}\label{e:sph_kernel}
\end{align}
with $\vec{r}$ a $D$-dimensional distance vector and $\norm{\cdot}$ the L2-norm, i.e., $\norm{\vec{r}} = \left(\sum_{i=1}^{D} r_i^2\right)^{1/2}$.
We note that derivatives of these functions can be easily evaluated, as the derivative operator can be transferred to the kernel, for which analytic results can be computed in advance. Thus,
\begin{align}
  \grad A\left(\vec{x}\right)
   & = \int\df{\primed{\vec{x}}} A\left(\primed{\vec{x}}\right) \grad_{\vec{x}} W\paren*{\vec{x} - \primed{\vec{x}}, h} \notag \\
  \grad \cdot \vec{V}\paren*{\vec{x}}
   & = \int \df{\primed{\vec{x}}} \vec{V}\paren{\primed{\vec{x}}} \cdot\grad_{\vec{x}}
  W\paren{\vec{x} - \primed{\vec{x}}, h} \notag\\
\end{align}

Numerically, these integrals are replaced by sums over the fluid particles, i.e., the interpolating points, such that
\begin{align}
  A(\vec{x}) = \int                                         & \df{\primed{\vec{x}}} A(\primed{\vec{x}}) W(\vec{x} - \primed{\vec{x}}) \notag                                \\
  \longrightarrow\psum_{\primed{j}\in \mathcal{R}(\vec{x})} & \frac{m_\primed{j}}{\rho_\primed{j}} A(\vec{x}_{\primed{j}}) W(\vec{x} - \vec{x}_\primed{j})\label{e:sph_sum}
\end{align}
with $m_i$ and $\rho_i$ the mass and density of the $i$-th fluid particle. To reduce the computational burden, these sums have been truncated to only include fluid particles within a cutoff region $\mathcal{R}(\vec{x})$ centered at $\vec{x}$, such that $\norm{\vec{x}_\primed{j} - \vec{x}} \le R_c = 3h$. For notational simplicity, in what follows we will denote these truncated sums using a primed summation symbol. As an example, the density of each element can be computed by taking $A=\rho$ and evaluating the function at $\vec{x}=\vec{x}_i$, with $A_i \equiv A(\vec{x}_i)$ and $W_{ij} = W(\vec{x}_i - \vec{x}_j)$
\begin{align}
  \rho_i = \psum_j m_j W_{ij}\label{e:sph_rhoi}
\end{align}
with the mass a constant.

To solve the equations of motion for the fluid particles, Eqs.~\eqref{e:sph_x}-\eqref{e:sph_v}, we need to evaluate the forces acting on each particle, which involves computing the divergence of the stress tensor. Instead of directly discretizing $\vec{\nabla}\cdot\tensor{\sigma}$, the ``golden-rules" of SPH state that formulas should be rewritten to place the density inside the differential operators\cite{1992}. In this way, the forces are evaluated using the following expression
\begin{align}
  \inv{\rho}\grad\cdot\paren*{\tensor{\sigma} - p\tensor{I}} & \equiv \grad\cdot\paren*{\frac{\tensor{\sigma} - p\tensor{I}}{\rho}} + \frac{\tensor{\sigma} - p\tensor{I}}{\rho^2}\grad\rho\label{e:sph_divstress}
\end{align}
Therefore, the time derivative of the fluid particle velocity $\vec{v}_i$, Eq.~\eqref{e:sph_v}, is given by
\begin{align}
  \dd{\vec{v}_i}{t} & =
  \sum_j m_j \bracket*{\paren*{\frac{\tensor{\sigma} - p \tensor{I}}{\rho^2}}_i + \paren*{\frac{\tensor{\sigma} - p \tensor{I}}{\rho^2}}_j}\cdot \grad_i W_{ij}\notag \\
                    & \quad + \vec{F}(\vec{x}_i)\label{e:sph_dvdt}
\end{align}
Finally, the equations of motion for the SPH particles are discretized in time and integrated using the following second-order scheme
\begin{align}
  \vec{x}_{i}^{n+1} & = \vec{x}_{i}^{n} + \vec{v}_{i}^{n}\Delta t + \frac{1}{2m_i}\left(\dd{\vec{v}_i}{t}\right)^{n}\left(\Delta t\right)^2\label{e:sph_int} \\
  \vec{v}_{i}^{n+1} & = \vec{v}_{i}^{n} +
  \frac{1}{2m_i}\left[\left(\dd{\vec{v}_i}{t}\right)^{n+1} + \left(\dd{\vec{v}_i}{t}\right)^{n} \right]\Delta t\notag
\end{align}

\subsection{Modified SPH}
Rigid boundaries, be they fixed walls or moving particles, can be incorporated within the SPH framework by discretizing them with boundary or wall particles.
However, using the standard SPH method introduced above, an unnatural flow is observed in the presence of such boundaries. To remove these artifacts, the values of the first and second derivatives of the physical quantities of interest can be used within the weighted averages\cite{Zhang2004a}, to arrive at the so-called Modified SPH (MSPH).
The appropriate expressions can be obtained by starting from the second-order Taylor expansion of the quantity of interest. For a scalar quantity $A\left(\vec{x}_j\right)$ we have
\begin{align}
  A(\vec{x}_j) & \simeq A(\vec{x}_i) + \grad A(\vec{x}_i)\cdot \paren*{\vec{x}_j - \vec{x}_i} \label{e:sph_taylor}                            \\
               & +\frac{1}{2} \transp{\paren*{\vec{x}_j-\vec{x}_i}}\cdot \grad_i\grad_i A(\vec{x}_i) \cdot \paren*{\vec{x}_j-\vec{x}_i}\notag \\
  A_j          & \simeq A_i + \grad A_i\cdot \vec{x}_{ji} + \frac{1}{2}\transp{\vec{x}}_{ji}\cdot\grad\grad A_i \cdot \vec{x}_{ji}\notag
\end{align}
with $\vec{x}_{ij}=\vec{x}_i-\vec{x}_j$. Multiplying both sides of Eq.~\eqref{e:sph_taylor} by the volume element $m_j/\rho_j$ times the smoothing function $W_{ij}$ and summing over all particles, we obtain
\begin{align}
  \sumsph{j} A_{j} W_{ij} & = \sumsph{j} W_{ij} A_i\label{e:msph_w}                                                     \\
                          & +\sumsph{j} W_{ij} \vec{x}_{ji}  \cdot \grad A_{i}\notag                                    \\
                          & + \frac{1}{2}\sumsph{j} W_{ij}\transp{\vec{x}}_{ji}\vec{x}_{ji}\colon\grad\grad A_{i}\notag
\end{align}
which relates $A$ to its first and second order derivates. To solve this equation, we then need the corresponding relations for $\grad A$ and $\grad\grad A$, which are given in Eqs.~\eqref{e:msph_dw}-\eqref{e:msph_ddw}
\begin{align}
  \sumsph{j} A_j \grad_i W_{ij} & = \sumsph{j} \grad_i W_{ij} A_i\label{e:msph_dw}                                                  \\
                                & + \sumsph{j} \grad_i W_{ij} \vec{x}_{ji}\cdot \grad A_i\notag                                     \\
                                & = \frac{1}{2}\sumsph{j} \grad W_{ij} \transp{\vec{x}_{ji}}\vec{x}_{ji}\colon\grad\grad A_i \notag
\end{align}
\begin{align}
  \sumsph{j} A_j \grad_i\grad_i W_{ij} & = \sumsph{j}\grad_i\grad_i W_{ij} A_i\label{e:msph_ddw}                                                   \\
                                       & + \sumsph{j}\grad_i\grad_i W_{ij} \vec{x}_{ji} \cdot\grad A_i \notag                                      \\
                                       & + \frac{1}{2}\sumsph{j} \grad_i\grad_i W_{ij} \transp{\vec{x}}_{ji}\vec{x}_{ji}\colon\grad\grad A_i\notag
\end{align}

Eqs.~\eqref{e:msph_w}-\eqref{e:msph_ddw} can be conveniently expressed in matrix form as follows:
\begin{align}
  \tensor{t}_{i}=\tensor{B}_{i}\cdot\tensor{f}_{i}\label{e:msph_tBf}
\end{align}
where $\tensor{f}_i = \transp{\begin{pmatrix} A_i, & \grad A_i, & \grad\grad A_i\end{pmatrix}}$ is a vector whose entries are formed from $A_i$ and its derivatives. In 2D this results in six independent components, thanks to the commutativity of the partial derivatives, such that
\begin{align}
  \tensor{f}_i & \equiv\begin{pmatrix}
     & A_{i\phantom{,xx}} \\
     & A_{i,x\phantom{x}} \\
     & A_{i,y\phantom{x}} \\
     & A_{i,xx}           \\
     & A_{i,xy}           \\
     & A_{i,yy}
  \end{pmatrix}\label{e:msph_f}
\end{align}
where commas are used to denote partial derivatives, $\partial_\alpha A_i= A_{i,\alpha}$ and $\partial_\alpha\partial_\beta A_i = A_{i,\alpha\beta}$. Likewise, the vector $\tensor{t}_i$ is composed using $W_{ij}$ and it's derivatives
\begin{align}
  \stensor{t}_{i}^{K} & = \sumsph{j}A_j \stensor{\Phi}_{ij}^{K}\label{e:msph_t}
\end{align}
where $\stensor{\Phi}_{ij}^{K}$ $(K=1,\ldots,6)$ are the components of $\tensor{\Phi}_{ij} = \transp{\paren*{W_{ij}, \grad_i W_{ij}, \grad_i\grad_i W_{ij}}}$, given by
\begin{align}
  \tensor{\Phi}_{ij} & \equiv\begin{pmatrix}
    W_{ij\phantom{,xx}} \\
    W_{ij,x\phantom{x}} \\
    W_{ij,y\phantom{x}} \\
    W_{ij,xx}           \\
    W_{ij,xy}           \\
    W_{ij,yy}
  \end{pmatrix}\label{e:msph_phi}
\end{align}
Finally, the $\tensor{B}_{i}$ matrix is defined as
\begin{align}
  \stensor{B}_{i}^{KL} & = \sumsph{j} \stensor{\Phi}_{ij}^K \stensor{\Theta}_{ij}^L\label{e:msph_B}
\end{align}
with $\stensor{\Theta}_{ij}^L$ the components of the $\tensor{\Theta}_{ij}$ vector, given by
\begin{align}
  \tensor{\Theta}_{ij} & = \begin{pmatrix}
     & 1                                \\
     & \paren*{x_i-x_j}                 \\
     & \paren*{y_i-y_j}                 \\
     & \frac{1}{2}\paren*{x_i-x_j}^2    \\
     & \paren*{x_i-x_j}\paren*{y_i-y_j} \\
     & \frac{1}{2}\paren*{y_i-y_j}^2
  \end{pmatrix}\label{e:msph_theta}
\end{align}
Within the MSPH method, we use $\tensor{f}_i = \tensor{B}_i^{-1}\cdot\tensor{t}_i$ to define the physical quantity $A_i$ of particle $i$ at position $\vec{x}_i$. Similar expressions can be defined for vector quantities and higher order tensors.

\subsection{Boundary Conditions}
To enforce the no-slip boundary condition at the fluid/solid interface, we use a virtual particle method\cite{Adami2012,Sato2019a}. The virtual particles are placed by reflecting the outermost-layer of wall particles with respect to the boundary. Then, the velocity vector at the positions of these virtual particles $\vec{v}_i^{\text{(virtual)}}$ is computed using the MSPH method, and the velocities of the symmetric wall particles are set according to $\vec{v}_i^{\text{(wall)}} = -\vec{v}_i^{\text{(virtual)}}$. In this way, the weighted average of the particle velocities at the boundary is guaranteed to be zero.
\begin{acknowledgements}
  The authors would like to thank Prof. Ryoichi Yamamoto, Prof. Matthew Turner and Dr. Simon K. Schnyder for fruitful discussion. This work was supported by the Japan Society for the Promotion of Science (Grants-in-Aid for Scientific Research KAKENHI no. 19H01862 and Wakate B no. 17K17825), the Ogasawara Foundation, and the SPIRITS 2020 of Kyoto University. Figures and movies were generated using Matplotlib\cite{Hunter2007}, a Python 2D plotting library.
\end{acknowledgements}
\clearpage
%\bibliography{gpmss}
%
\end{document}